\documentclass[useAMS,usenatbib]{mn2e}
\usepackage{graphicx}
\usepackage{epstopdf}
\usepackage{boxedminipage}
\usepackage{float}
\usepackage{url}

\title{Fossil evidence for spin alignment of SDSS galaxies in filaments}

\author[Bernard J.T. Jones et al.]{\parbox{\textwidth}{  Bernard J.T. Jones$^1$, Rien van de Weygaert$^1$ and Miguel A.~Arag\'on-Calvo$^2$,}
\vspace*{4pt}\\
$^1$Kapteyn Astronomical Institute, University of Groningen, P.O. Box 800, 9700 AV Groningen, The Netherlands \\
$^2$The Johns Hopkins University, Baltimore, Maryland, USA; miguel@pha.jhu.edu}

\begin{document}

% --------------------------------
% !!Remove what is in the braces!!
%\date{Source file version 1.4 Jan 22nd 2010}
% --------------------------------
\pagerange{\pageref{firstpage}--\pageref{lastpage}} \pubyear{2009}

\maketitle

\label{firstpage}

\begin{abstract}
We search for and find fossil evidence that the distribution of the spin axes of galaxies in cosmic web filaments relative to their host filaments are not randomly distributed.  This would indicate that the action of large scale tidal torques effected the alignments of galaxies located in cosmic filaments.  

To this end, we constructed a catalogue of clean filaments containing edge-on galaxies.  We started by applying the Multiscale Morphology Filter (MMF) technique to the galaxies in a redshift-distortion corrected version of the Sloan Digital Sky Survey DR5.  From that sample we extracted those 426 filaments that contained edge-on galaxies ($ b/a < 0.2$). These filaments were then visually classified relative to a variety of quality criteria.  These selected filaments contained 69 edge-on galaxies.  Statistical analysis using ``feature measures'' indicates that the distribution of orientations of these edge-on galaxies relative to their parent filament deviate significantly from what would be expected on the basis of a random distribution of orientations.  Fewer than $1\%$ of orientation histograms generated from simulated random distributions show the same features as observed in the data histogram.

The interpretation of this result may not be immediately apparent, but it is easy to identify a population of 14 objects whose spin axes are aligned perpendicular to the spine of the parent filament ($\cos \theta < 0.2$).  The candidate objects are found in relatively less dense filaments.  This might be expected since galaxies in such locations suffer less interaction with surrounding galaxies, and consequently better preserve their tidally induced orientations relative to the parent filament.  These objects are also less intrinsically bright and smaller than their counterparts elsewhere in the filaments.

The technique of searching for fossil evidence of alignment yields relatively few candidate objects, but it does not suffer from the dilution effects inherent in correlation analysis of large samples.  The candidate objects could be the subjects of a program of observations aimed at understanding in what way they might differ from their non-aligned counterparts.
\end{abstract}


% =========================================================
\section{Introduction}
\label{sec1:intro}
% ---------------------------------------------------------
We are searching for observational evidence that galaxies are aligned with the large scale structure in which they are embedded.  Searches for large scale galaxy alignments have a long and chequered history going back many decades. It is only with the advent of the great galaxy redshift catalogues, 2dFRS and SDSS, of accurate systematic galaxy photometry and of effective techniques for segmentation of the galaxy distribution into voids, filaments and clusters that we can now confidently address this phenomenon.  The recent review by \citet{Schafer08} covers many aspects of the subject of galaxy alignments from both the theoretical and observational points of view.

In this paper we specifically search for direct evidence for the alignment of the angular momentum of galaxies in the SDSS (DR5) catalogue 
of galaxies with the filamentary features in which they are embedded. 

Our approach in this paper is novel and so we briefly review what has been achieved using other, more standard, methods.  We also review the theoretical motivation driving our approach: it is our assertion that much of the large scale systemic alignment (as opposed to mere pairwise correlations) is driven by the tidal fields of the large scale structure.  Much of that motivation is based on large scale numerical simulations.

\subsection{The Cosmic Web}
%.............................
Large galaxy surveys such as the 2dF \citep{Colless01} and the Sloan Digital Sky Survey \citep{York00} have unambiguously revealed an intricate network of galaxies: the cosmic web.  The cosmic web can be described as a mixture of three basic morphologies: blob-like dense and compact \textit{clusters}, elongated \textit{filaments} of galaxies and large planar \textit{walls} which delineate vast empty regions referred to as \textit{voids} \citep{Zeldovich70, Shandarin89}.  \cite{Bondweb96} emphasized that this web-like pattern is shaped by the large scale tidal force fields whose source is the inhomogeneous matter distribution itself (see also \cite{Weybond09}). Within the large scale matter distribution, filaments are characterized 
by a strong and coherent tidal force field: they are tidal bridges in between massive cluster nodes. 

\subsection{Galaxy Alignments in the Cosmic Web}
\label{sec:alignweb}
With the large scale tidal force field responsible for the morphology and structure of the Cosmic Web, we expect an intimate link between 
the Cosmic Web and other prominent tidal manifestations of the structure formation process. According to the tidal torque theory tidal 
forces generate the angular momentum of collapsing halos. Thus we would expect the angular momentum of cosmic haloes to be correlated with 
the features of the cosmic web in which they are embedded. 

Such correlations would arise from the local tidal influence of those structures, though they would also be destroyed by local dynamical processes 
such as merging and orbital mixing \citep{Coutts96}. This suggests filaments to be the best places to look for alignments: not only are they 
sites with a strong and coherent tidal force field, but galaxies in filaments are also likely to have a less disturbed history than galaxies 
in cluster environments.

% ---------------------------------------------------------
\subsubsection{Spin alignment relative to structure}
% ---------------------------------------------------------
Methods based on correlations between pairs of galaxies do not identify systemic large scale alignments or correlations of alignments with specific types of structures such as clusters or filaments: they merely tell us that the spin orientations are not random.  The main problem here is to accurately identify structures and the objects that they contain. \citet{Han_etal95} examined a cosmic filament of galaxies, but could find no evidence for any sort of alignment. 

Recent advances in cosmic web structure analysis have provided methods whereby galaxy redshift catalogues can objectively be segmented into voids, filament and clusters. A first step in this direction was taken by \citet{Trujillo06} who, using the SDSS and 2dFRS catalogues, found evidence that spiral 
galaxies located on the shells of the largest cosmic voids have rotation axes that lie preferentially on the void surface. Their result tied in with 
numerical expectations of such an alignment by \cite{Brunino07}. 

% ---------------------------------------------------------
\subsubsection{The Sloan Digital Sky Survey and 2MASS}
% ---------------------------------------------------------
The Sloan Digital Sky Survey (SDSS) has opened up many possibilities for analysing the distribution of galaxy orientations on a variety of scales.  Local, small scale ($< 0.5h^{-1}$Mpc), correlations among galaxy orientations have been reported by \citet{SlosarWhite09}, \citet{Slosar09} and \citet{Jimenez10}, while \citet{Brainerd_etal09} and \citet{Cervantes_etal10} performed an extensive correlation analysis of the relative alignments of pairs of galaxies in the DR7 release of the SDSS.  The latter authors detected a significant correlation out to projected distance of 10 $h^{-1}$ Mpc and were even able to show differences with galaxy brightness.  The result compared well with the same analysis on simulated galaxy catalogues created from the Millennium Run Simulation. \cite{Paz08} correlated galaxy and halo shape with large scale structure and also claimed evidence for a significant correlation in both their numerical models and in the SDSS (DR-6).  \citet{Land08} examined the evidence for a violation of large-scale statistical isotropy in the distribution of projected spin vectors of spiral galaxies, using data on line of sight spin direction classified by members of the public partaking in the online project ``Galaxy Zoo''.  They found no evidence for overall preferred handedness of the Universe, even though the related study of spin correlations in Galaxy Zoo by \citet{Slosar09} and \citet{Jimenez10} did find a hint of a small-scale alignment.

\citet{LeeErdog07} determined tidal fields from the 2MASS redshift survey (2MRS) and looked at the orientation of galaxies in the Tully Galaxy Catalogue relative to this tidal field.  They found clear evidence of correlation  between the galaxy spins and the intermediate principal axes of the tidal shear (see appendix~\ref{app:tides}). This ties in with orientation studies based on galaxy catalogues which indicate that the spin vector of galaxies tends to be aligned with the plane of the wall in which the halo is embedded \citep{Kashikawa92, Navarro04, Trujillo06}.  

Paradoxically, there are no systematic studies of the orientation of galaxies in filaments despite their relatively easier identification.  We plan to use the edge-on galaxies in a filament catalogue in a marked correlation analysis.

% ---------------------------------------------------------
\subsubsection{Previous history}
% ---------------------------------------------------------
The earliest commentaries on this subject (\citet{Brown38} and \citet{Reynolds22} citing even earlier work by Brown) pre-date the establishment of an extragalactic distance scale by Hubble.  Brown reported the detection of a significant systemic orientation of spiral galaxies having small inclinations to the line of sight.  Much later, after the publication of the Palomar Sky Survey, Brown published two further papers on the subject \citep{Brown64,Brown68} based on visual examination of thousands of galaxies covering vast areas of sky, reporting evidence for significant alignment.  \citet{Reaves58}, using specially prepared plate material, was unable to confirm Brown's claims. 

Brown's work was later taken up by \citet{HawleyPeebles75} who removed possible systemic biases by using a double blind experiment.  They searched for evidence of overall alignment, alignment between neighbouring pairs and alignments in the Coma cluster of galaxies.  None of these searches yielded statistically significant evidence for galaxy alignment.

But that was far from the end of the story.  The Local Supercluster (LSC) was analysed by \citet{JaanisteSaar78}, \citet{MacGillivray82} and \citet{FlinGodlowski86} with contradictory results, and with no significant evidence emerging.  However, the latest attempts at resolving  the question of alignments in the Local Supercluster are those of \citet{Hu_etal06} and \citet{Ayral08a, Ayral08b} in which it is claimed that there is evidence of significant alignment of some galaxy sub-populations, but not others.  The situation from these surveys is manifestly unclear.  The best that can be said is that, even if there is any systemic alignment, it is far from easy to find evidence for or to quantify the phenomenon using approaches based on analysing samples having large numbers of galaxies.   

% ---------------------------------------------------------
\subsection{A clean filament catalogue}
% ---------------------------------------------------------
In this paper we try an alternative approach in which we build a small but statistically useful catalogue of filaments containing edge-on galaxies.   The use of edge-on galaxies simplifies the problem of determining the space orientation of the galaxies.  The filaments are places where, a priori, we might expect to find an alignment effect.  Our catalogue is not in any way biased with respect to the orientations of the galaxies defining those filaments.

The plan of the paper is to review the theoretical notions underlying spin orientation correlations with large scale structure, and comment on what N-Body experiments tell us in this regard.  Then we go on to discuss how, using the DR5 release of the SDSS, we build a small catalogue of cosmic web filaments that are``good places to seek out orientation anomalies''.    

% ---------------------------------------------------------
\subsection{Outline}
% ---------------------------------------------------------
We start by discussing the context of our study in section~\ref{sec:theory}, the expected alignment of the angular momentum with the 
surrounding weblike structures. Subsequently, we present our SDSS DR5 galaxy sample in section~\ref{sec:sdss}, along with the applied corrections for artefacts such as redshift distortions. Section~\ref{sec:filament} elaborates on the construction of our FILCAT-0 filament catalogue, and the selection of the class A filaments which will be subjected to our feature analysis. On the basis of the geometry of the galaxy-filament alignment configuration, in 
section~\ref{sec:filgeom} we will derive a statistical distribution function for the random galaxy-filament orientation distribution which forms the reference model for our study. The galaxy orientations of edge-on galaxies in our selection of SDSS filaments is analysed in section~\ref{sec:data_analysis}, followed by the statistical feature measure analysis in section ~\ref{sec:statmodel}. Armed with a firmly established significant alignment of a small subset of 14 galaxies, in section~\ref{sec:objects} we address the question of the nature of these aligned objects. Finally, section~\ref{sec:conclusions} summarizes and discusses our results.

\section{Tidal Manifestations: \\ \ \ \ \ \ Alignments and the Cosmic Web}
\label{sec:theory}
%.............................
Within the context of gravitational instability, the gravitational tidal forces 
establish an intricate relationship between some of the most prominent manifestations 
of the structure formation process (for a proper analytical definition of the tidal 
shear field $T_{ij}$ we refer to appendix~\ref{app:tides}).

\subsubsection{the Cosmic Web}
Perhaps the most prominent manifestation of the tidal shear forces is that of the distinct 
{\it weblike geometry} of the cosmic matter distribution. The Megaparsec scale tidal shear forces 
are the main agent for the contraction of matter into the sheets and filaments which trace out the 
cosmic web. The main source of this tidal force field are the compact dense and massive cluster 
peaks. This is the reason behind the strong link between filaments and cluster peaks: filaments 
should be seen as tidal bridges between cluster peaks \citep{Bondweb96,Weyedb1996,Weybond09}.

\subsubsection{Angular Momentum and Tidal Torques}
\label{sec:torque}
In addition to the filamentary cosmic web itself, we recognize the manifestations of the large-scale 
tidal fields over a range of scales. Perhaps the most important influence is their role in generating the 
angular momentum and rotation of dark haloes and galaxies. It is now generally accepted that galaxies derive 
their angular momentum as a consequence of the action of tidal torques induced by the surrounding matter 
distribution, either from neighbouring proto-galaxies or from the large scale structure in which they 
are embedded. 

As a result of the tidal force field, a collapsing halo will get torqued into a rotating object. The magnitude 
and direction of the resulting angular momentum vector $L_i$ of a protogalaxy is related to the inertia tensor, 
$I_{ij}$, of the torqued object and the driving tidal forces described by the tidal tensor $T_{ij}$ (\ref{eq:quadtide}),
\begin{equation}
L_i \propto \epsilon_{ijk} T_{jm} I_{mk},
\label{eq:angmom}
\end{equation}
with $\epsilon_{ijk}$ the Levi-Civita symbol, and where summation is implied over the repeated indices \citep{White84}.

The idea that galaxy spin originated through tidal torques originated with \citet{Hoyle49} 
who suggested that the source of tides was the cluster in which the galaxy was embedded.  This was put into the 
context of modern structure formation by \citet{Peebles69} and \citet{Dorosh1970}, where 
\citet{Peebles69} focussed on the tidal interactions between neighbouring proto-galaxies as being 
the source of the rotation. 

However, confusion concerning the efficiency of the mechanism remained widespread until the numerical study 
by \citet{EfJon79} and many generations of simulations thereafter \citep[e.g.][]{Barnes87, Dubin1993, Porciani02, 
Knebe08}. Although the early simulations did not attempt to identify the source of the tidal field, numerical 
simulations along with related analytical studies demonstrated the viability of the tidal torque 
mechanism \citep{White84,Ryden1987,HeavPeac1988,Catelan1996,LeePen01,Porciani02}. 

\subsubsection{Aligning Spin and Web}
\label{sec:align}
The tidally induced rotation of galaxies implies a link between the surrounding external matter 
and the galaxy formation process itself. With the cosmic web as a direct manifestation of the large 
scale tidal field, we may therefore wonder whether we can detect a connection with the angular 
momentum of galaxies or galaxy halos. The theoretical studies of \citet{Sugerman2000} and \citet{LeePen00} 
were important in pointing out that this connection should be visible in the orientation of 
galaxy spins with the surrounding large scale structure. 

The notion that galaxy spins might be correlated with large scale structure has been extensively discussed on the 
basis of data catalogues, even those that existed in the early days of large scale studies 
(see sect.~\ref{sec:alignweb}). Theoretically, the interpretation of this alignment is facilitated 
by invoking the parameterized formalism forwarded by \cite{LeePen00} to describe the correlation 
between the angular momentum ${\bf L}$ (eqn.~\ref{eq:angmom}) and the tidal tensor $T_{ij}$. To this 
end, they introduced the parameter $c$ to quantify the autocorrelation tensor of the angular momentum 
in a given tidal field, 
\begin{equation}
\langle L_i L_j | {\bf T} \rangle \propto {1 \over 3} \delta_{ij} + c ({1 \over 3} \delta_{ij} - T_{ik}T_{kj} )
\end{equation}
where $c = 0$ corresponds to the situation where tidal and inertia tensor would be mutually independent and 
the resulting angular momentum vectors would be randomly distributed. In appendix~\ref{app:tidalign} we summarize 
the main ingredients of this description. 

The alignment of galaxy haloes, the spins of galaxies and dark halos, of clusters and of voids with 
the surrounding large scale structures have been the subject of numerous studies \citep[see e.g.][]{Binggeli1982,RheeKatgert87,
Plionis02,Plionis03,Basilakos06,Trujillo06,Altay06,Aragon07_spins,Hahn07a,Leevrard07,LeeSpringel08,Platen08}. The expectation of 
a strong correlation between the orientation of structures and the large scale force field configuration is based on the 
existence of such correlations in the primordial Gaussian density field \citep{Bond1987,Desjacques08}. Conversely, one should  
recognize the important role assigned to the alignment of rich clusters in determining the strength and morphology 
of the cosmic web \citep{Bondweb96}.

% ---------------------------------------------------------
%\subsubsection{Evidence from N-Body simulations}
% ---------------------------------------------------------
\subsubsection{Aligning Spin and Web:\\
\ \ \ \ \ \ \ \ \ evidence from N-body simulations}
\label{sec:nbodyalign}
N-Body simulations provide information on the alignment of the orientations of galaxy halos and galaxy spins. 
With this it is possible to discuss the correlations of spins among galaxies and the correlations of spin with 
large scale structures such as filaments and walls. The main difficulty in the latter case is to unambiguously 
identify those large scale structures. The recent advances in cosmic web structure analysis have enabled 
a systematic segmentation of the cosmic matter distribution into voids, sheets, filament and clusters.

As a result, recent $N$-body simulations \citep{Aragon07_spins, Hahn07a, Hahn07b, Paz08, Hahn09, 
HahnPhd09,Zhang09} have been able to find, amongst others, that the filamentary or sheetlike nature of the 
environment has a distinct influence on the shape and spin orientation of dark matter haloes. In the case 
of of haloes located in large scale walls, they seem to agree that both the spin vector and the major axis 
of inertia lie in the plane of the wall. In the case of the alignment of halos with their embedding filaments, 
\cite{Aragon07_spins} and \cite{Hahn07a} found evidence for a mass and redshift dependence, which has 
been confirmed by the studies of \cite{Paz08} and \cite{Zhang09}. This mass segregation involves a 
parallel alignment of spin vector with the filament if the mass of the halo is less than the 
characteristic halo mass $m < 10^{12}$ M$_{\odot}$ h$^{-1}$, turning into a perpendicular 
orientation for more massive haloes \citep[also see][]{Bailin04}. A tantalizing related finding is that 
by \citet{Faltenbacher09}, who claims correlations between spin and large scale structure extending over 
scales of over $100h^{-1}$Mpc. 

Results from $N$-body simulations can not be directly compared with observations. The reason for this is that the only available 
information of the spin of galaxies comes from the luminous baryonic matter. The coupling between the spin of the baryonic and 
dark matter components is still not well understood. One of the key issues raised by studies of simulations is that the dark matter 
halos and the baryonic disk component may not line up with one another. Recent high-resolution hydrodynamical simulations of the formation 
of galaxies within a cosmic filament by \cite{HahnPhd09, Hahn10} confirms, or even strengthens, the earlier trend observed by \citet{Bosch02} 
and \citet{Chen03}, who found that a median angle between gas and dark matter is in the order of $\simeq 30^{\circ}$ \citep[also see][]{Bett09} 
This misalignment between dark matter and gas is sufficient to erase any primordial alignment signal between the galaxies and their host 
filament or wall. This effect however, could be rendered negligible if the gaseous component of galaxies retains its primordial 
orientation better than its dark matter counterpart as suggested by the SPH simulation study of \citet{Navarro04}. The filament simulation 
by \cite{HahnPhd09} \citep[also see][]{Hahn10} seems to be considerably less optimistic in this respect. 

% =========================================================
%     Data sample
% ---------------------------------------------------------
\section{The SDSS Galaxy sample}
\label{sec:sdss}
% ---------------------------------------------------------

\begin{figure*}
  \vskip 0.5truecm
  \centering
  \includegraphics[width=\textwidth]{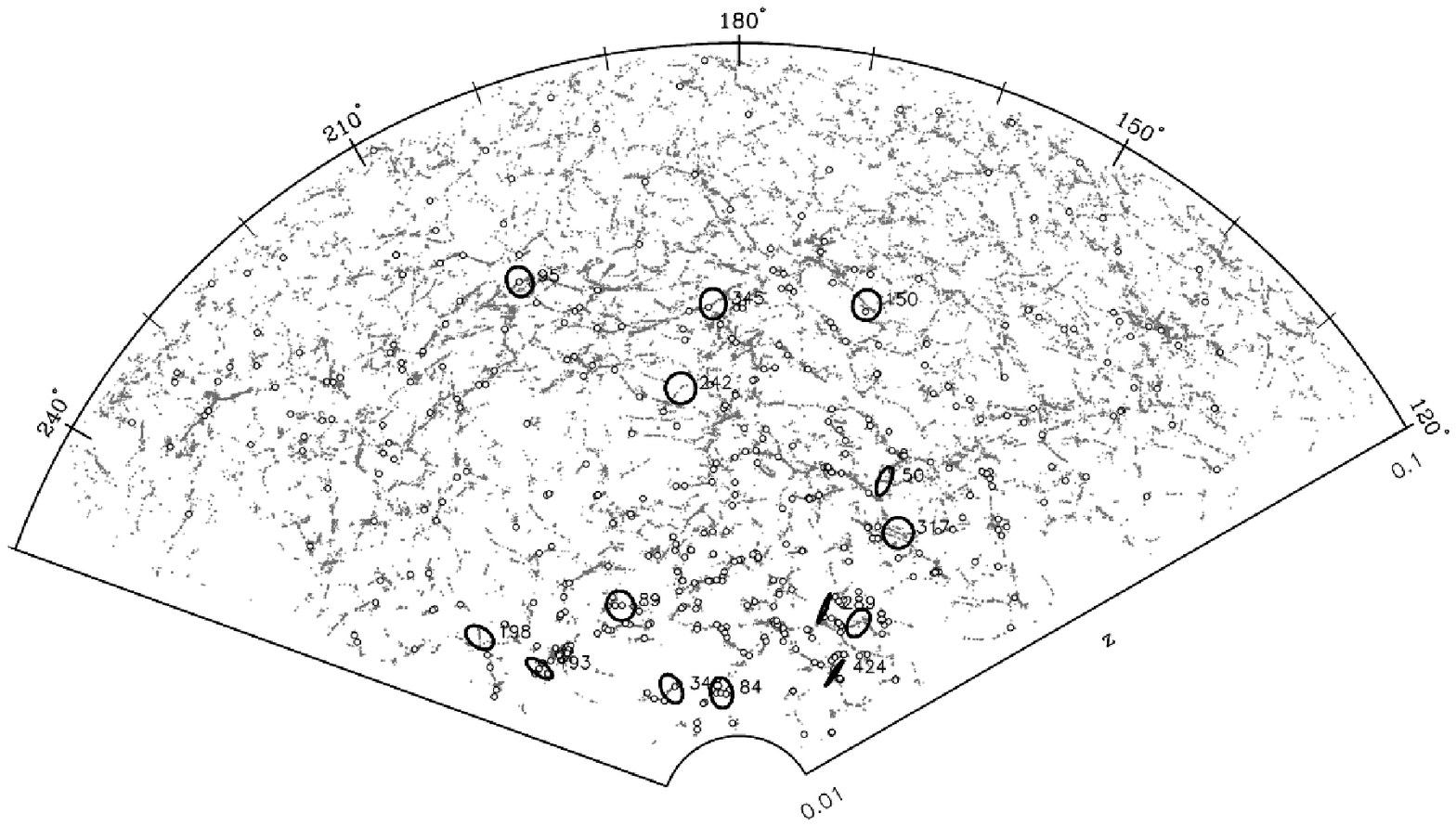} 
  \includegraphics[width=\textwidth]{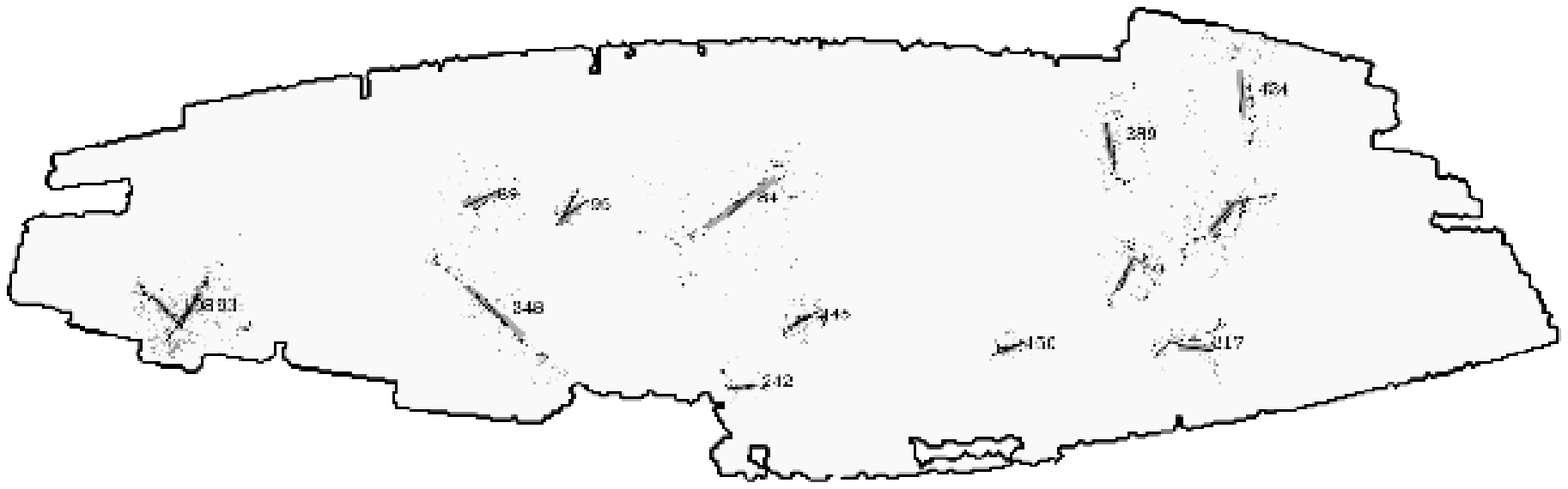} 
  \caption{The SDSS DR5 sample volume seen in two projections. Top: redshift slice through the volume, out to $z=0.1$. 
The image shows the SDSS galaxies in the sample volume that have been identified by MMF as filament galaxies. The galaxies are shown at the location assigned by the MMF filament compression algorithm (see Arag\'on-Calvo et al. 2007). The small open circles indicate the location of all 492 edge-on galaxies in our sample. The 14 large ellipses, and corresponding number, locates the 14 edge-on galaxies that our analysis identifies as significantly aligned with their parent filament. The elliptical shape corresponds to the orientation of a circle with a similar orientation as the galaxy. Bottom: sky region of the SDSS DR5 survey, including the location of the 14 positively aligned galaxies (see fig.~19 for galaxy images). The edge-on galaxies are plotted as lines indicating the galaxies' sky orientation. Each of these galaxies is shown against the backdrop of the projection of their embedding filament.}
  \label{fig:SDSSsurvey} 
\vskip 0.5truecm
\end{figure*} 

The results presented in this letter are based on a galaxy sample selected from the largest galaxy survey to date, the Sloan Digital Sky Survey data release 5 (SDSS DR5).  The SDSS is a wide-field photometric and spectroscopic survey carried out with a dedicated 2.5 meter telescope at Apache Point, New Mexico \citep{York00}. The telescope scans continuously the sky on five photometric band-passes namely $u$, $g$, $r$, $i$ and $z$, down to a limiting $r$-band magnitude of 22.5 \citep{Fukugita96, Smith02}. 

The analysis presented in this letter includes approximately $100,000$ SDSS DR5 galaxies located in the range 
\begin{eqnarray}
&&0<\eta<40\nonumber\\
&&0.01<z<0.11\nonumber
\end{eqnarray}
within the \emph{survey coordinate system} ($\lambda$, $\eta$) \citep{Stoughton02}. This geometry encloses a volume of approximately  
$7.3\times10^6$ Mpc$^3$h$^{-3}$.

In order to take full advantage of the galaxies in the sample we used a magnitude limited catalogue. This means that the radial distribution of galaxies must be weighted in order to produce an isotropic distribution.  We used a simple formula to model the change in the mean number of galaxies as function of their redshift given by \citet{Efstathiou01}: 
\begin{equation}
dN = A z^2 \exp{( -(z/z_r)^\beta ) }.
\end{equation}
Where $A$ is a normalization factor that depends on the density of galaxies, $z_r$ is the characteristic redshift of the distribution and $\beta$ encodes the slope of the curve.

Before applying filament finding algorithms, like MMF, it is necessary to fill in any holes in the survey (see section \ref{subsec:unsampled}) and handle the effects of the boundaries of the surveyed region (see section \ref{subsec:boundaries}).

% =========================================================
\subsection{Data systematics, removal of artefacts and the final filament samples}
% ---------------------------------------------------------
The catalogued galaxy distribution is affected by several systematics and artefacts. Among the most important are telescope artefacts, unsampled areas of the sky, peculiar velocities from both linear and non-linear processes.

% ---------------------------------------------------------
\subsubsection{Fingers of God}
\label{subsec:fogfilaments}
% ---------------------------------------------------------
Fingers of God have a characteristic elongated shape that will introduce ``false'' filaments in the distribution of galaxies along the line of sight. In order to correct this effect we compressed the fingers of God in a similar way as done by \citet{Tegmark04} by identifying elongated structures and them compressing them along the line of sight. One drawback of the finger of God compression algorithm is that it systematically removes filaments oriented along the line of sight. This is a consequence of those filaments being confused with fingers of God. In our analysis we take account of this.

% ---------------------------------------------------------
\subsubsection{Unsampled regions}
\label{subsec:unsampled}
% ---------------------------------------------------------
We correct the holes in the angular mask by means of a new method that exploits the volume filling properties of the Delaunay tessellation \citep{Aragon07_thesis}.  The Delaunay Tessellation Field Interpolator is a geometry-based interpolation scheme that uses the spatial information at the edges of the hole and follows the tessellation to derive the density field in the interior of the hole.

% --------------------------------------------------------
\subsubsection{Boundaries}
\label{subsec:boundaries}
% ---------------------------------------------------------
In order to produce an uniform density field around the survey volume we place particles, randomly, with the same radial distribution as the galaxies in the sample, outside the sample volume. 

\subsubsection{Edge-on selection}
Optical surveys such as the SDSS provide enough information to derive the position angle, inclination and, in some cases, the sense of rotation of the galaxy. However we cannot determine which is the approaching side or the receding side.  This results in a four-fold degeneracy in the derived spin vector. The study of spin alignments only requires the direction of the spin vector and not its sense of rotation. This reduces the degeneracy to two-fold which is still insufficient. 

There are two special cases where the spin vector can be unambiguously determined only from the position angle and inclination: \textbf{i}) Edge-on galaxies having their spin vector in the plane of the sky. \textbf{ii}) Face-on galaxies having their spin vector pointing along the line-of-sight.

In this paper we restrict our analysis to edge-on galaxies.  Edge-on galaxies are identified in terns of the ratio of their projected axes by $r_b/r_a < 0.2$.  Strictly speaking, this applies only to spiral galaxies which are assumed to be flat rotating disks with their spin vectors pointing perpendicular to the disk.  (We could use colour information to separate early and late type galaxies if that were necessary, but we do not do that since the flattening of the objects in our sample is extreme).  Using the axial ratio criterion we can measure the inclination of face-on and edge-on galaxies with an uncertainty of $\pm 12^{\circ}$.  Imposing a stricter selection criterion on the axial ratio would improve this, but it would also decrease the size of the useful sample of galaxies.

Within the boundaries of our SDSS DR5 subsample we found 492 edge-on galaxies. 

% ---------------------------------------------------------
\section{Filament Catalogue:\\ \ \ \ \ \ Detection and Selection}
\label{sec:filament}
% ---------------------------------------------------------
%--------------------------------------------------------------------------
%     Finding and classifying structure
%--------------------------------------------------------------------------
Even if there were systemic, non-random, orientations set up by, say, global tidal fields at the time of galaxy formation, subsequent dynamical evolution would conspire to largely eradicate the phenomenon from the present data.  Here we think of interactions between galaxies among themselves or with neighbouring large scale structures such as clusters.  Our approach, then, is to look for fossils that have not undergone such evolution and so retain a vestige of their original orientation.

To accomplish this we use techniques that are perhaps more familiar in data mining: we cull and clean the dataset to obtain a subset of the very best data that is available, and then analyse the result to look for potentially interesting anomalies in the distribution of orientations. We address the general significance of context of this strategy below, in sect.~\ref{sec:kdd}.

We use a scale free structure finding algorithm (MMF, see below) to identify cosmic web filaments, restricting ourselves to the edge-on galaxies within those filaments (see above).  The galaxy spin axis orientation is then in the plane of the sky, and since MMF yields the space orientation of the host filament, we can calculate for the entire sample the distribution of spin axis - filament angles.  The goal is then to answer the question as to whether these orientations are random, and if not, say whether there are preferred orientations. 

\begin{figure*}
  \centering
  \includegraphics[width=\textwidth]{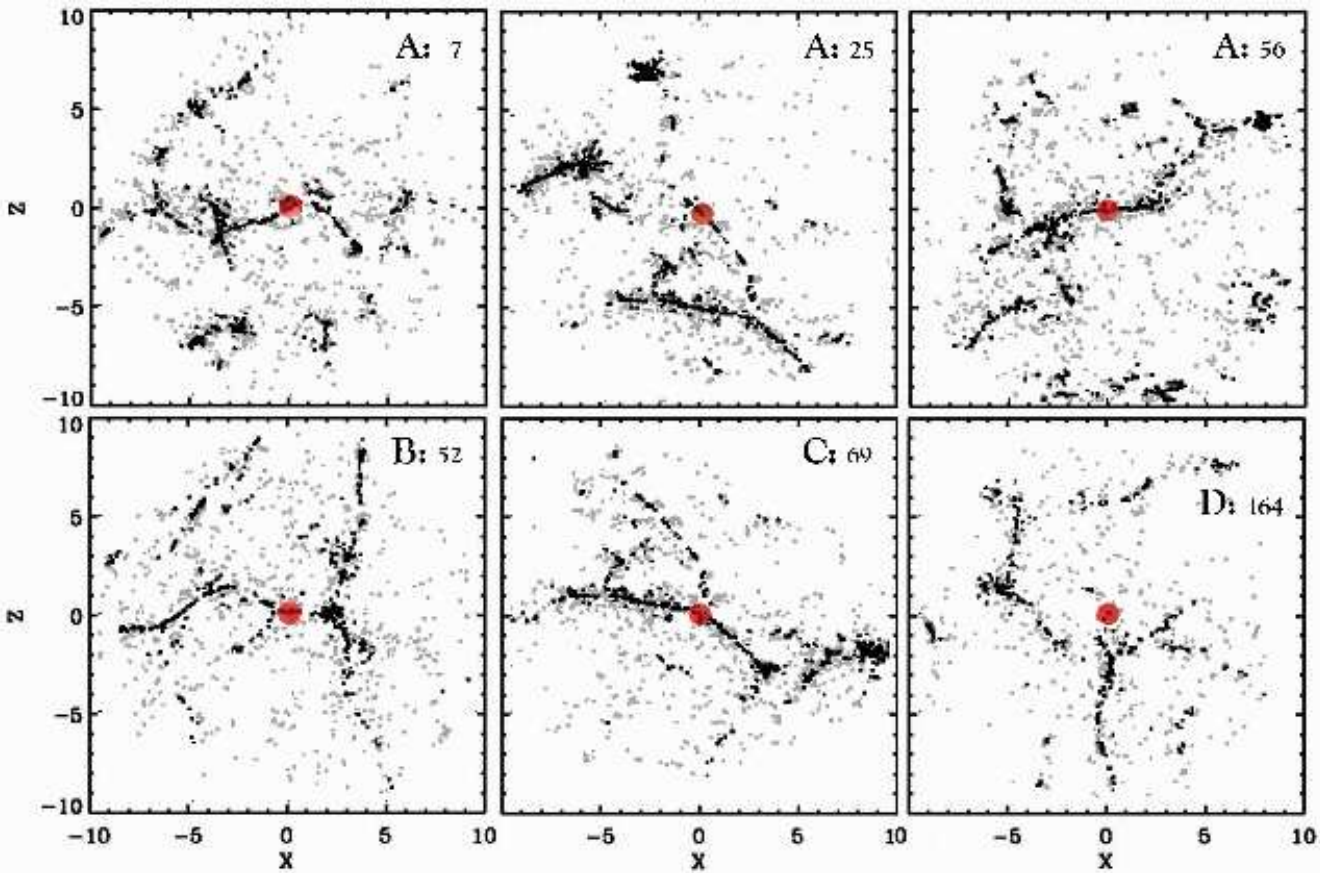} 
  \caption{A selection of 6 filaments showing the compressed filament and the location of the host galaxy.  Each panel shows the identification of the filament in the FILCAT-0 filament list and the assigned classification. Black dots: galaxy location after filament compression (Arag\'on-Calvo 2007). Grey dots: original SDSS location.}
  \label{fig:Classification} 
\end{figure*} 
\subsection{Sample analysis: KDD data mining} 
\label{sec:kdd}
The current approach differs substantially from previous attempts to find evidence for alignments between galaxy spins and large scale structures in that we search, not for a trend or correlation in the data, but for specific candidate examples of such alignments. By focussing on edge-on galaxies in filaments, and then grading the resultant filaments we arrive at a relatively small database of high quality filaments hosting edge-on galaxies that can be analysed from the point of view of relative galaxy-filament orientation. To do this we follow the precepts of the so-called KDD process for analysing large, possibly heterogeneous or partially sampled, data sets. It involves several steps including culling and cleaning of the search data set, in this case the DR5 release of SDSS.  

KDD stands for ``Knowledge Discovery in Databases'', a term introduced by \citet{{Piatetsky91}}. KDD is described by \citet{{Fayyad96}} as ``The non-trivial process of identifying valid, novel, potentially useful, and ultimately understandable patterns in data''.  The more familiar concept of Data Mining is very much a part of KDD, rather than the other way around\footnote{There is a good lecture by Susan Imberman at \url{http://www.cs.csi.cuny.edu/~imberman/DataMining/KDD beginnings.pdf}}. 

An integral part of the KDD approach is to have a measure of ``interestingness'' that ranks what we are looking for.  Note that this is not necessarily the same as the more familiar concept of {\it significance}.  In this respect we study the histogram of the distribution of spin orientation of a galaxy relative to the parent structure, identifying features in the histogram that make it notable, or interesting, with respect to the problem at hand.  We introduce a ``Feature measure'' that reflects this level of interestingness.

Formally, the process of extracting a suitable feature measure for the present study is based on wavelet analysis of the histogram of the distributions of orientations from the data sample itself and from a simple simulation. 

\begin{figure*}
  \centering
  \includegraphics[width=\textwidth]{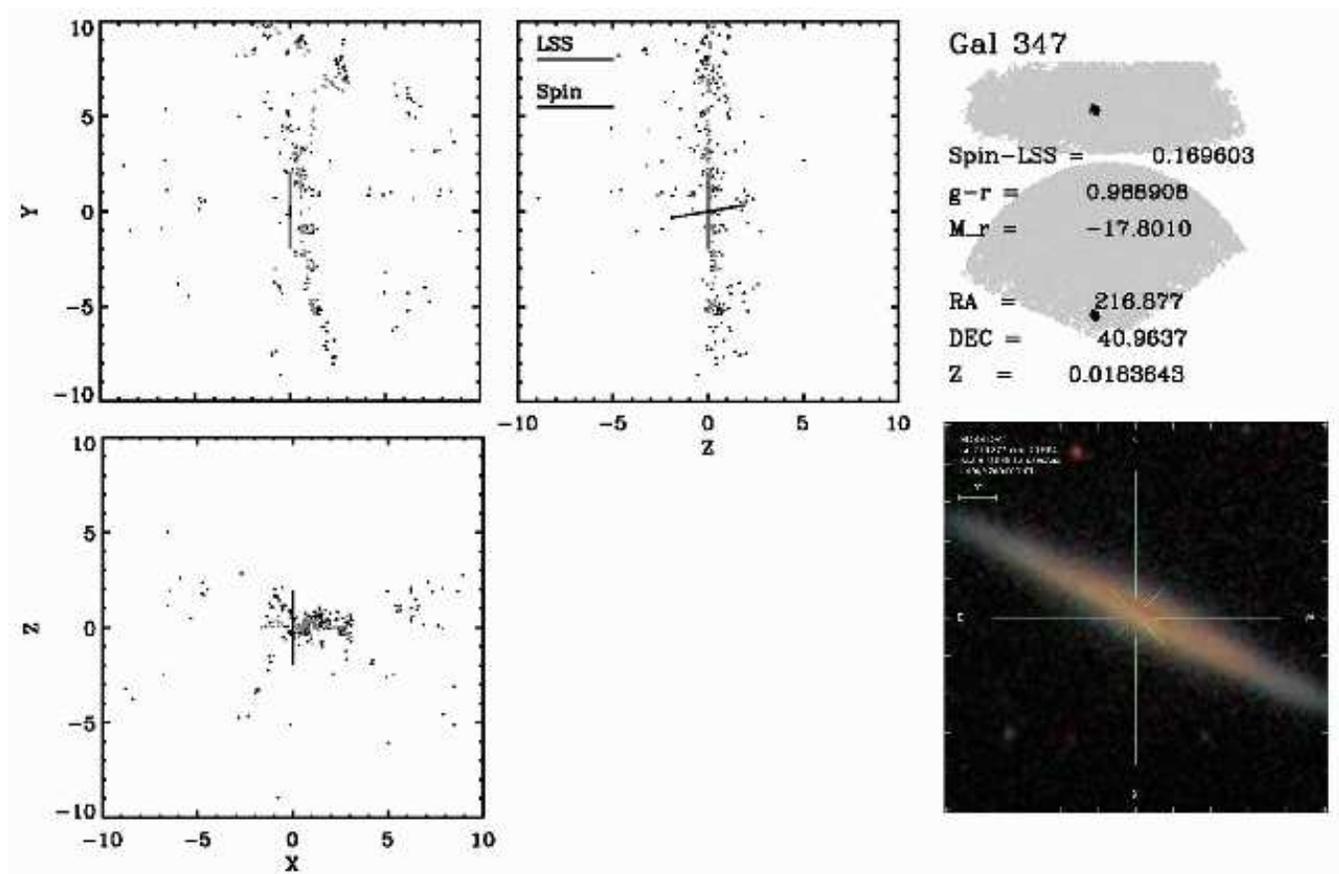} 
  \caption{Specific example of an edge-on sample galaxy, nr. 347 of our sample. Bottom right: SDSS  image of edge-on sample galaxy nr. 347. Top 
right: location of galaxy nr. 347 on the SDSS DR5 sky region and redshift region, along with some key data. The position and orientation of 
the galaxy with respect to its embedding filament is depicted by means of three mutually perpendicular squares of 20h$^{-1}$ Mpc size, 
centered on the galaxy. Black dots: galaxy location after filament compression (Arag\'on-Calvo 2007). Grey dots: original SDSS location.}
  \label{fig:TypicalGalaxy} 
\end{figure*} 
% ---------------------------------------------------------
\subsection{MMF Filament identification and classification}
% ---------------------------------------------------------
Revealing correlations between galaxy spin direction and the large scale structure requires the ability to unambiguously identify the morphological features of the Cosmic Web.  Several methods have been used in an attempt to identify and extract the morphological components of the Megaparsec-scale matter distribution \citep[among others]{Barrow85, Babul92, Luo95, Stoica05, Colberg05, Pimblet05,Sousbie08} with varying degrees of success. 

Key to our approach is to have a first class structure finder so that we can unambiguously identify the filaments and the member galaxies, and so find any edge-on galaxies they might be hosting. It should be stressed that in any study of galaxy orientation relative to parent structure, it is important to identify that parent structure as accurately as possible. The results presented here are based on the Multiscale Morphology Filter MMF \citep{Aragon07_MMF}. Based on a density field reconstructed from SDSS using the DTFE process \citep{Schaapwey00, Weyschaap09}, MMF largely achieves the goal of outlining the filamentary spine of the Cosmic Web. However, as we shall see, it is worth culling the filament sample to provide a filament sample that is most suited to this task.  We want to avoid dilution of what, in the light of past studies, is evidently at best a weak effect.

The morphological characterisation achieved by MMF enables us to isolate specific host morphological environments  for galaxies and test predictions from the tidal torque theory in a systematic way. 

% ---------------------------------------------------------
\subsubsection{The Multiscale Morphology Filter}
% ---------------------------------------------------------
The MMF objectively segments a point set representing the cosmic web into its three basic morphological components: clusters, planes and filaments.  It does this first by using the DTFE methodology to create a continuous density field from the point set, and then hierarchically analysing the properties of the local matter distribution.  The DTFE method produces an optimal reconstruction of the continuous density field, retaining the characteristic hierarchical and anisotropic nature of the Cosmic Web. This allows MMF, multi-scale by construction, to produce a catalogue of filaments with relatively few scale biases.  

The Multiscale Morphology Filter is based on the second-order local variations of the density field as encoded in the Hessian matrix ($\partial^2 \rho / \partial x_i \partial x_j$).  For a given set of smoothing scales we compute the eigenvalues of the Hessian matrix at each position on the density field.  We use a set of morphology filters based on relations between the eigenvalues in order to get a measure of local spherical symmetry, filamentariness or planarity. The morphological segmentation is performed in order of increasing degrees of freedom in the eigenvalues for each morphology (i.e. blobs $\rightarrow$ filaments). 

The response from the morphology filters computed at all scales is integrated into a single multi-scale response which encodes the morphological information present in the density field. In a filament, the eigenvectors of the Hessian matrix corresponding to the smallest eigenvalue  {$\mathbf{e_{F}}$) indicate the direction of filament.  Eigenvectors are computed from a smoothed version  of the density field in order to avoid small-scale variations in the direction assigned to filaments.

A little more detail is presented in Appendix \ref{app:mmf} for the sake of completeness.

% ---------------------------------------------------------
\subsubsection{The raw FILCAT-0 filament catalogue}
% ---------------------------------------------------------
On the basis of our MMF technology we construct a raw filament catalogue.  Before this can be used it is necessary to correct for the effects of redshift distortions and to go through a filament thinning (or compression) process to enhance the visual appearance for quality assessment (see sections  \ref{subsec:fogfilaments}).  The final stage is to grade filaments and select those deemed most useful for angular momentum fossil searches (see section \ref{subsec:classfilaments}).

The MMF method yields a sample of 426 filaments containing edge-on galaxies that we shall use in the analysis of the spin-filament relationship.  We refer to this catalogue as the FILCAT-0 filament catalogue \citep{Aragon07_thesis}.  Each filament can be viewed in three orthogonal projections, and for each filament there is a raw view (redshift space) and a processed view in which a correction for redshift distortion has been applied, followed by a compression of the member galaxies towards the spine of the filament.

Provided the selection takes no account of the orientation of the edge-on galaxy this procedure should be relatively benign. Accordingly, the classification was made in such a way that it is independent of the orientation of the sample galaxies, and also does not involve prior knowledge of the orientation of the filaments with respect to the line of sight. 

% --------------------------------------------------------
\subsection{Filament classification}
\label{subsec:classfilaments}
% ---------------------------------------------------------
Since MMF simply classifies structure on the basis of shape, the FILCAT-0 catalogue inevitably contains a variety of structures, all having a general linear structure.  In the forensic spirit of this investigation we have to pare this list down to a subset of filaments where we can reasonably expect to find evidence for any putative relationship between the edge-on galaxy and its host filament.  

Filaments were scored 1, 2, or 3 depending on the visual appearance of the filament on three orthogonal projections of the redshift distortion corrected and compressed filament.  Scoring was done independently by BJ and RvdW, and the scores totalled.  Filaments scoring '1' by both, i.e. having a total score of 2, were assigned to ``Class A'', those with a total score of `3' to ``Class B'', those with score `4'  to ``Class C'' and the rest to ``Class D''.

The 1-2-3 classification of filaments depended on a subjective quality assessment of the filament and on the location of the edge-on galaxy relative to the filament.  The following criteria were used for filament quality:
\begin{enumerate}
\item 
 must be defined by a reasonable number of galaxies
\item 
 must not be too sparse
\item 
 must not have a branched structure
\item 
 must be clearly defined in all three projections
\end{enumerate}
This reflects the use to which the filament is to be put: we are looking for places where we might reasonably expect the local tidal dynamics to be simple.  In keeping with that approach we further require that the key edge-on galaxy: 
\begin{enumerate}
\item 

 must be located on or near the filament
\item 
 must not lie in a sparse area of the filament
\item 
 must not lie at the end of the filament
\end{enumerate}
The procedure in applying these criteria is as follows.  For a ``1'' score the filament must be clearly defined in all three planes, and it must have a good population of points.  The logic is that if it is not seen in all three planes, it might be a wall.  Thus if this criterion fails in one plane it is ranked as a ``3''.  If the filament is well defined but sparse it is given a ``2''.  Finally ``1'' and ``2'' filaments are downgrade to a ``3'' if the galaxy is not on a filament (subjective!) or if it sitting at the end or junction of filaments (a Y-shape).   This is not unlike the old-fashioned 1950's and 1960's approach to classifying objects like clusters of galaxies.

Although the orientation relative to the line of sight was not a classification factor, it turned out that all filaments making an angle closer than about $40^{0}$ with the line of sight were eliminated.  This is shown in figure \ref{fig:lineofsight} where the distribution of angles the filaments make with the line of sight is shown for the entire sample and for the Class A sub-sample.

The final list of Class A filaments contained 69 objects only from the original sample of 492 edge-on galaxies. There are 106 Class A and Class B filaments, containing in total 121 edge-on galaxies.

\begin{figure}
  \centering
  \includegraphics[width=\columnwidth]{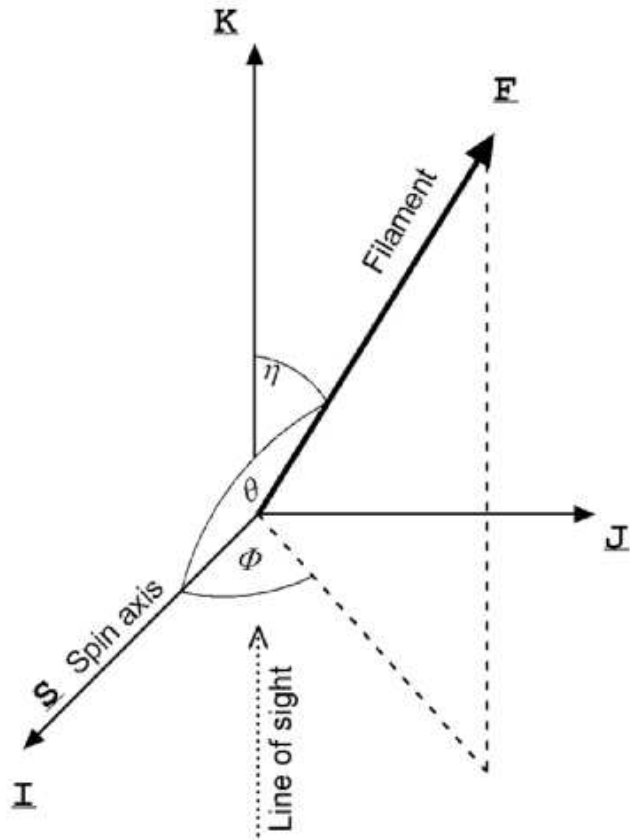} 
  \caption{Vectors and angles defining the orientation of the filament to the line of sight and the orientation of the galaxy minor axis on the plane of the sky.  The line of sight looks along the $\vec{K}$-axis and the galaxy minor axis defines the $\vec{I}$-axis in the p[lane of the sky.  The vector $\vec{J}$ completes the axis triad relative to which the filament orientation is measured. }
  \label{fig:angles} 
\end{figure} 

% --------------------------------------------------------
\section{Geometry of Filament and Galaxy orientations}
\label{sec:filgeom}
% --------------------------------------------------------
\subsection{Angles}
%........................................................
Consider one edge-on galaxy in our sample. The spin axis of the galaxy (taken to be the minor axis of its shape) and the line of sight define a right handed coordinate system as shown in figure \ref{fig:angles}.  The filament will make an angle $\eta$ with the line of sight.  The projected angle between the galaxy spin axis and the projection of the filament on the plane of the sky will be denoted by $\Phi$.  The angle of interest for this study, $\theta$, is the angle between the filament and the galaxy spin axis.

There is a four-fold degeneracy in the definition of these angles.  Since the galaxy is only viewed edge on the spin axis $\vec{S}$ lies in the plane of the sky.  There is no additional information about the sense of rotation.  The spin axis is defined without a sense of direction and so we should be able to make the transformation $\vec{S} \rightarrow -\vec{S}$ without changing the angle $\theta$.  The line of the filament, $\vec{F}$ is also defined without a sense of direction and so making the transformation $\vec{F} \rightarrow -\vec{F}$ should not change the value assigned to $\theta$.

We can handle this degeneracy by restricting the ranges of the angles to $0 \le \eta \le \pi/2$ and $0 \le \Phi \le \pi$, and by forcing the range of $\theta$ to be $0 \le \theta \le \pi/2$.

The angle $\theta$ between the filament and the galaxy spin axis is defined in terms of the observables $\eta$ and $\Phi$ by
\begin{equation}
\cos \theta = |\sin \eta \cos \Phi|
\label{eq:thetadefinition}
\end{equation}
The absolute value is needed because of the degeneracies discussed above and to bring $\theta$ into the range $0 \le \theta \le \pi/2$.

It is the statistical distribution of the angle $\theta$ that concerns us.  We are comparing the distribution of this angle, or specifically the distribution of $\cos(\theta)$, with what would be expected if the spin axis and the spine of the filament direction were uncorrelated.  So, for each galaxy and its local filament direction, we compute $\cos(\theta)$.

%........................................................
\subsection{Uncorrelated spins and filaments} 

%........................................................
The angles $\eta$ and $\Phi$ are clearly independent.  In the absence of selection effects, $\cos\eta$ is uniformly distributed, as is the angle $\Phi$.  Note, however, that if there were a correlation between the directions of the spin axis and the filament, $\Phi$ would no longer be uniformly distributed. If the spin directions of galaxies are independent of the host filament, and the host filaments are themselves randomly oriented with respect to the line of sight, then the distribution of $\cos\theta$ will also be uniform.

However, owing to a variety of issues, the filament sample is not randomly oriented and so $\cos \eta$ is not uniformly distributed (see figure \ref{fig:lineofsight}).  Consequently, $\cos \theta$ will not be uniformly distributed either.  We see in figure \ref{fig:lineofsight} (bottom) that subsamples A 
and B of filaments have an almost uniform distribution of $\cos\eta$, except for a cutoff at higher values of $\cos\eta$. For our sample, this 
corresponds to an exclusion of all filaments within an angle of $\eta_c \approx 35^{\circ}-40^{\circ}$ to the line of sight 
(see sect.~\ref{sec:filmangle}). 

\begin{figure}
\vskip -0.1truecm
  \mbox{\hskip -0.35truecm\includegraphics[width=1.1\columnwidth]{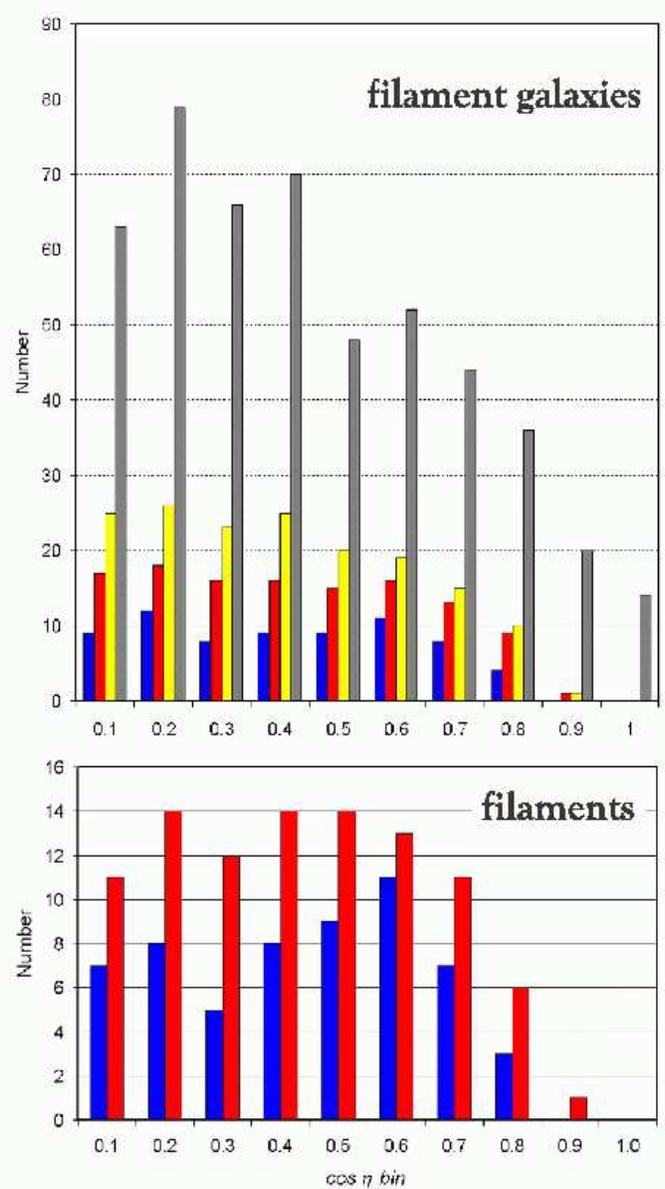}}
  \vskip -0.25truecm
  \caption{Distribution of the angle, $\eta$, the filaments in our sample make with the line of sight.  
Bottom: the $\cos \eta$ distribution of the 70 class A filament galaxies (solid red bars) and 121 class B filament galaxies 
(solid blue bars), along with the distribution of the edge-on filament galaxies embedded in these filaments (dashed bars). 
Top: the distribution of the $\cos \eta$ values of the embedding filaments of the entire sample sample of 492 edge-on filament 
galaxies is shown in grey along with three sub-samples: the 70 class A filament galaxies (dark blue) and the set of 121 Class B 
filament galaxies (red) and all edge-on galaxies in filaments not rated as '3' by either classifier (yellow).}
  \label{fig:lineofsight} 
\end{figure} 
If this is modeled with a sharp cutoff at angles $\eta < \eta_c$, the distribution of $\cos \theta$ can easily be derived in closed form.  The expected fraction of objects having a particular value of $\cos\theta$, relative to a uniform distribution is
\begin{eqnarray}
f &= \displaystyle \frac{2}{\pi} \arcsin \left[\frac{\cos \eta_c}{\sin \theta} \right], \quad &\theta > \pi /2 - \eta_c  \nonumber \\
 &= 1 & \textrm{otherwise}.
 \label{eq:eta_select}
\end{eqnarray} 
This is derived and compared with simulations in Appendix \ref{app:randomspins}.

Other factors come into assessing the distribution that would be seen even in the absence of a spin-filament correlation.  In particular, the sample of filaments found by MMF with the corrections for redshift distortion and the compression procedure produces a non-uniform distribution of $\cos(\eta)$ that is biased against including filaments lying along the line of sight.  

Subsequent visual selection of a sub-sample of filaments imposes additional constraints on the expected distribution.  As it turns out, these can be modelled and so we can compare the data with a well-reasoned model.

% --------------------------------------------------------
\section{Data analysis}
\label{sec:data_analysis}
%........................................................
\begin{figure*}
  \vskip -0.5truecm
%  \mbox{\hskip -1.0truecm\includegraphics[bb=0 0 1059 708,width=\textwidth]{figures_eps/ThetaClassA.png}} 
%  \mbox{\hskip -1.0truecm\includegraphics[bb=0 0 1059 708,width=\textwidth]{jonesspin.fig6.png}} 
  \mbox{\hskip -1.0truecm\includegraphics[width=\textwidth]{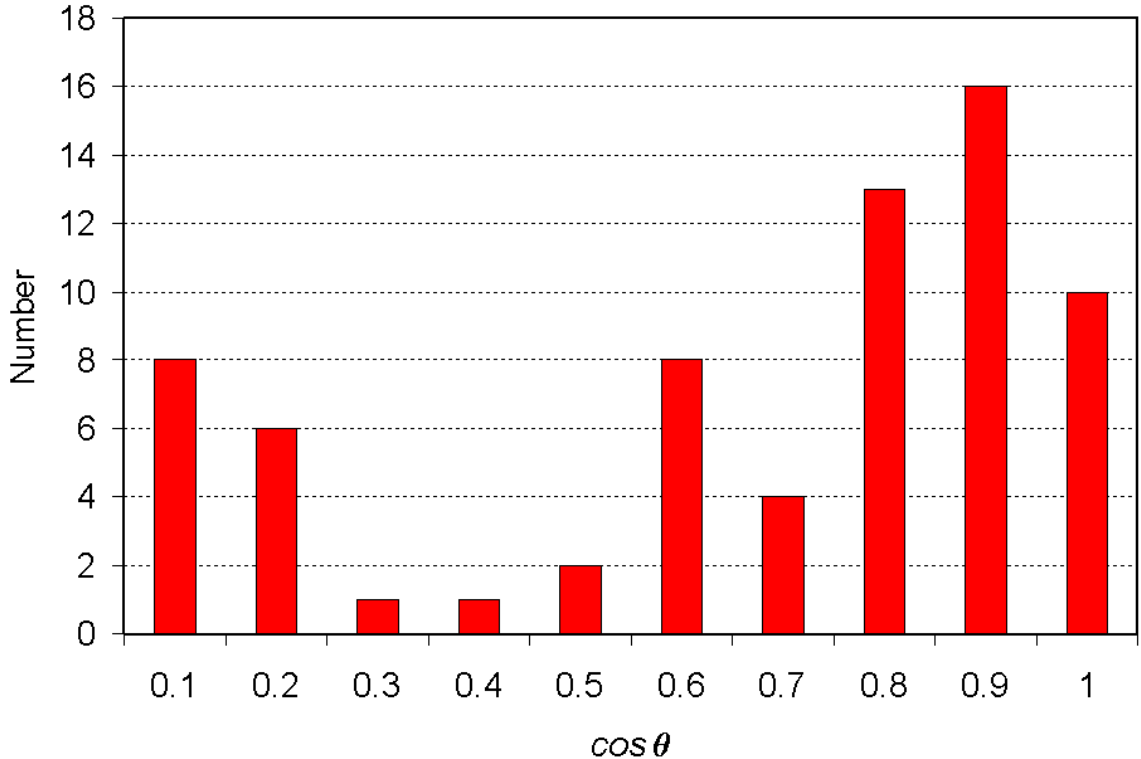}} 
  \vskip -0.5truecm
  \caption{Distribution of the angle $\theta$ the spin axis of the galaxy makes with the direction of its parent filament.  Two distributions are shown: one the distribution expected from the analytic result of equation (\ref{eq:thetadistribution}) and the other for the galaxies in the sample of Class A filaments.  The hump in the filament data at $\cos \theta < 0.2$ is outstanding.}
  \label{fig:dataangles} 
\end{figure*} 
\begin{figure}
  \includegraphics[width=\columnwidth]{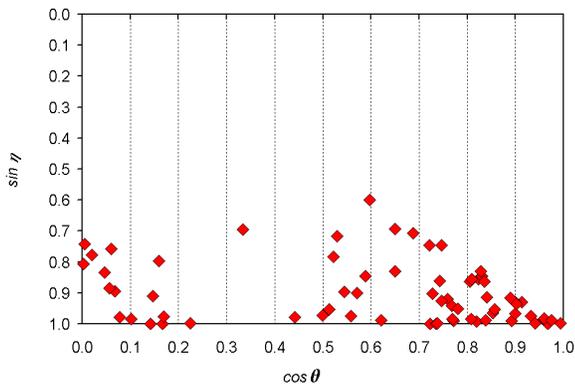} 
  \caption{Relationship between the inclination of the filament to the line of sight, $\eta$, and the relative orientation of the galaxy and the host filament $\theta$ for the Class A sample.  The horizontal lines delineate the histogram bins shown in figure \ref{fig:dataangles}.  The excess at low $\cos \theta$ is seen to be associated with filaments having $\sin \eta > 0.75$, i.e. tending to be transverse to the line of sight.  This should occasion no surprise as most of the Class A filaments turn out to be so-orientated.}
  \label{fig:Eta-Theta} 
\end{figure} 

We can derive for our sample of filaments, and for a variety of sub-samples, the angle that the filament makes with the line of sight.  This 
is shown in figure \ref{fig:lineofsight}.

In figure \ref{fig:lineofsight} we see the decline in the number of filaments as $\cos \eta \rightarrow 1$: the distribution for the 
entire sample is nowhere flat. It is clearly reflected in the angular distribution of parent filaments of the 492 edge-on filament 
galaxies (top panel). The drop-off at $\cos \eta > 0.8$ is hardly surprising in view of the overall distribution: it presumably reflects the 
prejudice against finding filaments lying along the line of sight. 

The distribution of $\cos \eta$ for the sub-samples of Class A and Class A+B filaments is uniform for $\cos \eta < 0.8$, as would be expected for 
a randomly oriented vector relative to the pole of the distribution (see lower panel, fig.~\ref{fig:lineofsight}, solid red and blue bars).  So, 
while the distribution of the line of sight angles for the whole sample is far from uniform, the Class A and selection criteria have led to 
a uniform distribution of angles that has been censored at angles less than around $35^{\circ}-40^{\circ}$ to the line of sight. Not only 
does this finding assure us that we understand the angular distribution of filaments (see sec.~\ref{app:randomspins}), but also provides 
a strong argument for the solidity of the MMF filament finding process. 

Because of this distribution of $\eta$, the distribution of $\cos \theta$, the angle the spin axis makes with the 
filament, is not uniform either. This readily follows from the geometric model described by equation~\ref{eq:eta_select}. 

\subsection{Filament angles with line of sight} 
\label{sec:filmangle}
%........................................................
%........................................................
\subsection{Orientations in our filament sample} 
%........................................................
First we should remark that the orientations of the spin axes relative to the sky are consistent with a uniform distribution: there is no bias towards specific orientations in the SDSS catalogue.  

The distribution of spin axis to filament angles, $\theta$, derived from our Class A sample is shown in figure \ref{fig:dataangles}.  We see a striking hump for objects with $\cos\theta < 0.2$: this contains 14 of the 82 galaxies while in an uncorrelated distribution we would have expected less than 1. 

It is on this basis that we draw two conclusions: 
\begin{enumerate}
\item 
The spin axes of these edge-on galaxies are correlated with the direction of their local filament in the sense that their spin axes are perpendicular to the filament spine. 
\item 
The filaments are of real physical significance (otherwise we would not see anything other than a uniform distribution, as per the non-selected data sample). 
\end{enumerate} 
 
It is important to take account of the fact that there are no filaments in our Class A sample making a small angle, $\eta <\pi / 4$, with the line of sight (see figure \ref{fig:lineofsight}).  Removing these from the sample further decreases the occupancy of the bins having spin-filament angles greater than $\pi/4$ (i.e. $\cos \theta < 0.7$) (see equation (\ref{eq:eta_select})).  The distribution of $\cos\theta$ for our Class A filament sample is shown in figure \ref{fig:dataangles}, and it is compared with the expectations of our simple uncorrelated spins distribution in figures \ref{fig:shuffletheta} and \ref{fig:simulatetheta}.

%--------------------------------------------------------
\subsection{The significance of the anomaly} 
% --------------------------------------------------------
The $\cos\theta < 0.2$ feature in the distribution is quite striking.  Simply considering the geometry as depicted in figure \ref{fig:angles} it is difficult to imagine any model based on random spin-filament orientation in which the number of objects increased with decreasing $\cos \theta$:  the available phase space for the angles $\eta$ and $\Phi$ does not allow that.

There are only 14 objects in this hump in the distribution, and so, having found it, the question arises as to why the other galaxies in the edge-on sample are not correlated in this way.  Why are these 14 galaxies so different from the rest ? 
 
Before addressing this we must however reassure ourselves that we have done nothing to the data that might give rise to this effect.  After all, we have created a very specific sub-sample from some 426 filaments hosting edge-on galaxies. 
 
Several factors have come into the sample definition: 
\begin{enumerate}
\item 
Corrections for "fingers of god" eliminate filaments along the line of sight.  Since for such a filament the spin axis will tend to be perpendicular to the filament, this elimination actually removes objects having small $\cos\theta$ from the sample!  So this does not help enhance our bump 
\item
We have rejected filaments in which the galaxy in question does not lie close to the spine of the filament.  Including such galaxies would only serve to randomise the distribution rather than enhance a feature at some particular angle.
\item 
We have rejected filaments for which a spine cannot unambiguously be defined.  This includes rejecting filaments that may be branching in or around the position of the edge-on galaxy. 
\end{enumerate} 

% ---------------------------------------------------------
\subsection{Other samples - a comparison}
% ---------------------------------------------------------
We shall, in the next section, make a statistical assessment of this result by comparing it with simulated samples.  It is useful in the first instance to look at a variety of other samples.  Firstly we shall randomise the spin vectors among the filaments in the Class A sample to show that this result disappears.  Then we shall comment on whether the entire catalogue shows any evidence of this, finally we shall look at the edge-on galaxies in filaments that appear to lie in the plane of the sky.

\begin{figure}
  \vskip -0.5truecm
%  \mbox{\hskip -1.25truecm\includegraphics[bb=0 0 1056 704,width=1.2\columnwidth]{figures_eps/AllFilaments_theta.png}} 
%  \mbox{\hskip -1.25truecm\includegraphics[bb=0 0 1056 704,width=1.2\columnwidth]{jonesspin.fig8.png}} 
  \mbox{\hskip -1.25truecm\includegraphics[width=1.2\columnwidth]{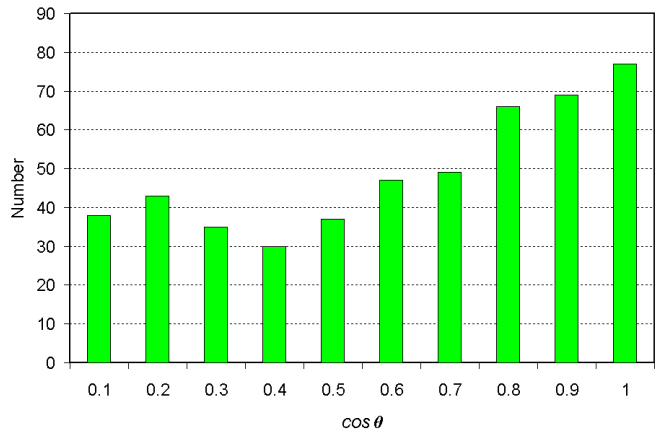}} 
  \vskip -0.25truecm
  \caption{Distribution of the orientation of the spin axis of an edge-on galaxy relative to the direction of the host filament, for the entire sample of filaments.}
  \label{fig:AllFilaments_theta} 
\end{figure} 

% ---------------------------------------------------------
\subsubsection{The entire sample}
% ---------------------------------------------------------
When classifying the filaments we rejected around 70\% of the original sample.  It is interesting to compare the selected Class A sub-sample with the original sample from which it was extracted, remembering of course that the original sample had a non-uniform distribution of filament orientations, and that the position of the edge-on galaxy was not constrained.

If we repeat the analysis for a selection of galaxies from poorly defined filaments we see a distribution that is consistent with the spins being uncorrelated.  This is shown in figure \ref{fig:AllFilaments_theta}.  The distribution is almost uniform, but there is nonetheless a discernible feature at low $\cos(\theta)$ which we might convince ourselves is the same hump as in the selected sub-sample.  However, unlike in the selected sample, this hump does not have a striking level of significance and by itself would not constitute particularly compelling evidence for alignment.
\begin{figure}
  \vskip -0.5truecm
%  \mbox{\hskip -1.2truecm\includegraphics[bb=0 0 1061 707,width=1.2\columnwidth]{figures_eps/SkyPlaneFilaments.png}} 
%  \mbox{\hskip -1.2truecm\includegraphics[bb=0 0 1061 707,width=1.2\columnwidth]{jonesspin.fig9.png}} 
  \mbox{\hskip -1.2truecm\includegraphics[width=1.2\columnwidth]{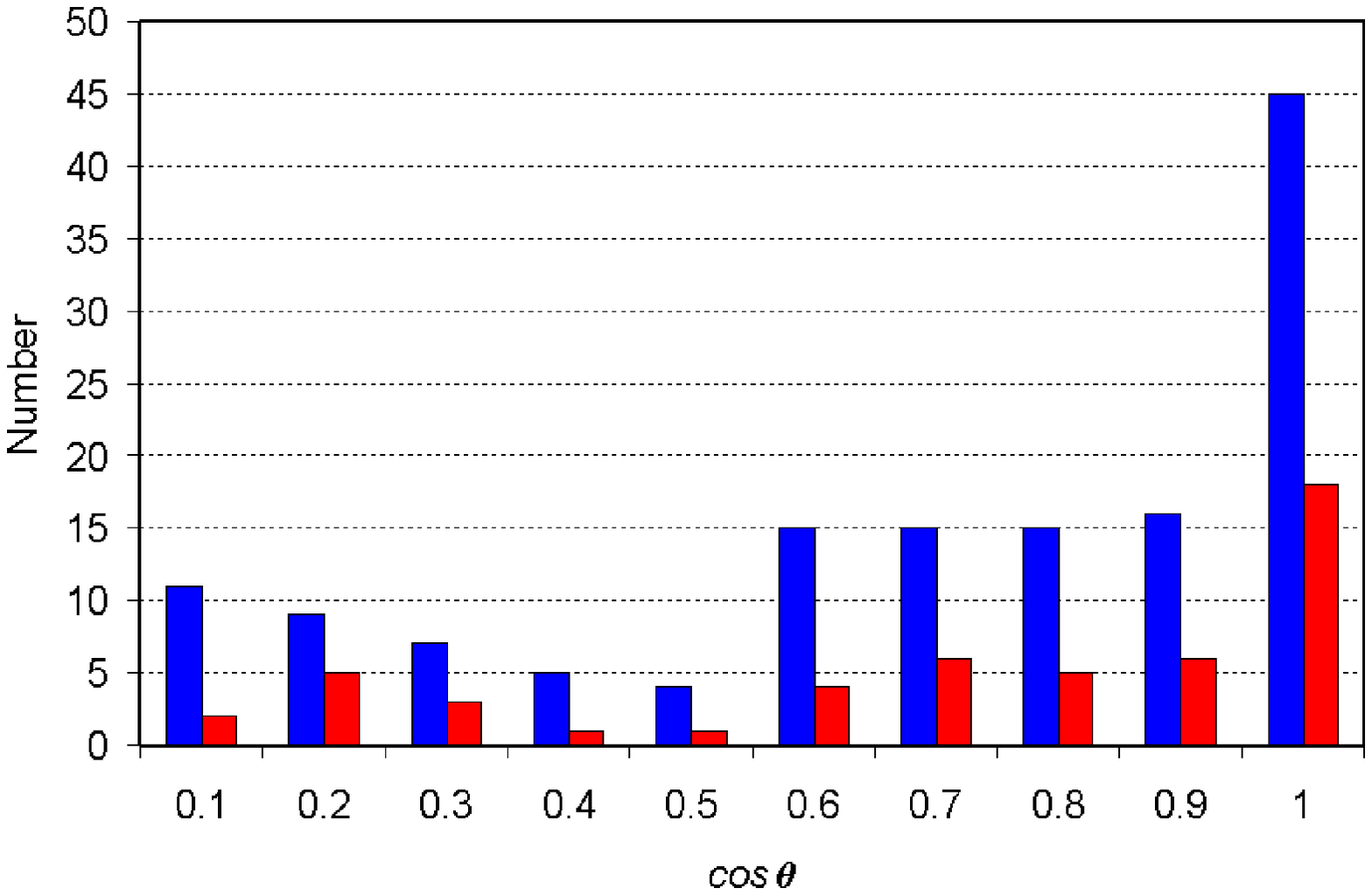}} 
  \vskip -0.25truecm
  \caption{Filaments in the plane of the sky: distribution of the angle $\theta$ the spin axis of the edge-on galaxies make with the host filaments.  The blue histogram is the complete sample of sky-plane SDSS filaments and the red histogram is a sub-sample excluding poorly rated filaments (see text).}
  \label{fig:skyplanetheta} 
%\end{figure} 
%\begin{figure}
%  \includegraphics[bb=0 0 790 566,width=\columnwidth]{figures_eps/ShuffleSpins.png} 
%  \includegraphics[bb=0 0 790 566,width=\columnwidth]{jonesspin.fig10.png} 
  \includegraphics[width=\columnwidth]{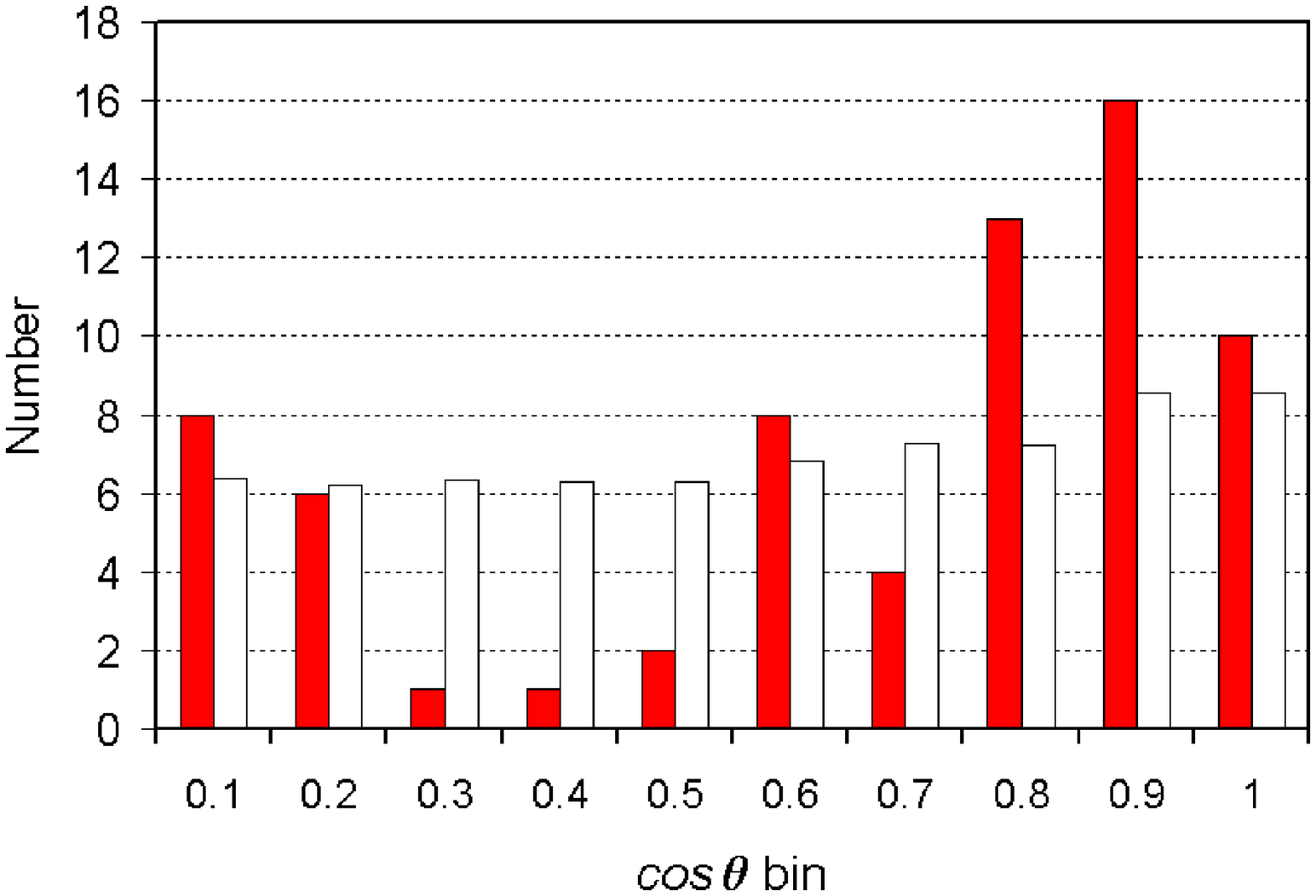} 
  \vskip -0.25truecm
  \caption{Red: distribution of the angle, $\theta$ in bins of equal $\cos\theta$ for the Class A sample of spin vectors.   White:  The expected distribution of spin axis directions when edge-on galaxies in our sample are assigned to randomised host filaments. }
  \label{fig:shuffletheta} 
%\end{figure} 
%\begin{figure}
%  \includegraphics[bb=0 0 790 566,width=\columnwidth]{figures_eps/DataWithModel.png} 
%  \includegraphics[bb=0 0 790 566,width=\columnwidth]{jonesspin.fig11.png} 
  \includegraphics[width=\columnwidth]{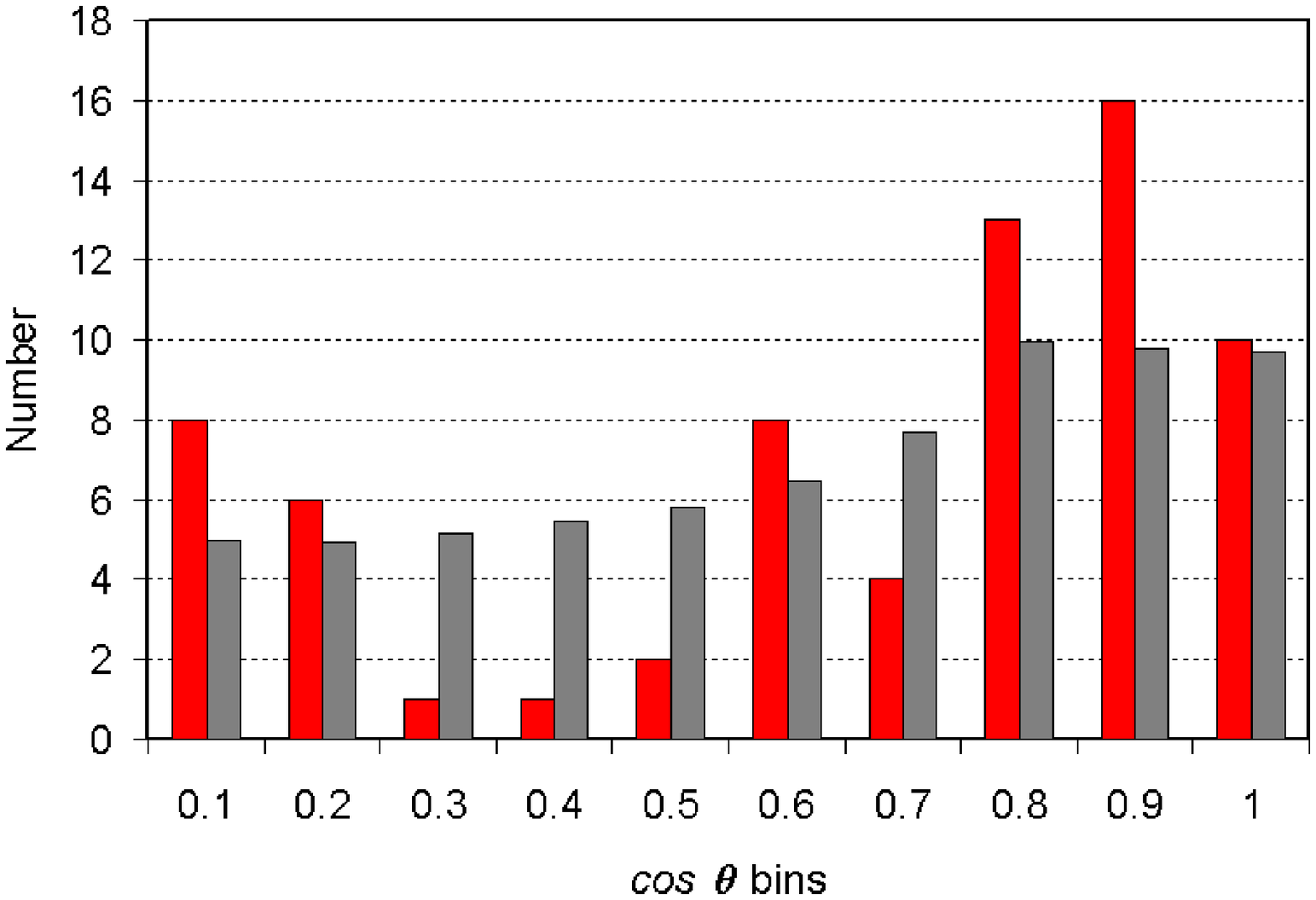} 
  \vskip -0.25truecm
  \caption{Red: distribution of the angle, $\theta$ in bins of equal $\cos\theta$ for the Class A sample of spin vectors.   Grey:  distribution $\theta$ in bins of equal $\cos\theta$ expected according to a model for the filament selection process consistent with figure \ref{fig:lineofsight}. }
  \label{fig:simulatetheta} 
\end{figure} 

%........................................................
\subsubsection{Filaments in the plane of the sky}
%........................................................
Filaments in the plane of the sky (or close to it) offer another opportunity to test for alignments within this sample.  In this case $\sin\eta \approx 1$ in equation \ref{eq:thetadefinition} and the angle between the filament and the spin axis is precisely what is seen.  For such filaments  the redshift distortion is presumably irrelevant, causing a broadening of the galaxy distribution. 

In order to maintain the quality of the sample we restrict ourselves to sky-plane filaments rated '1' or '2' by both BJ and RvdW: i.e. we reject any filament rated '3' by either.  This leaves a sample of 51 filaments.  The distribution of $\cos\theta$ for this sample is shown in figure \ref{fig:skyplanetheta}.
The sample is selected primarily on the basis of the angle $\eta$.  It is perhaps surprising that there are only three filaments from the Class A sample having $\cos\theta < 0.2$ in this sample of sky-plane filaments: the samples are effectively independent.  The fact that we see in the sky-plane filaments a hump at low $\cos\theta$ is support for the conclusion suggested on the basis of the more rigorously selected Class A sample.

%........................................................
\section{Statistical modelling}
\label{sec:statmodel}
%........................................................
The Class A sample consists of 67 filaments, of which we single out 14 for special attention.  If we extend this to include the Class B filaments the sample grows to 121 filaments.  Generating a simple model for random spin orientations is straightforward, but the question remains how to evaluate the credibility of the observed distribution.  The ``natural'' test would be to compare the data with the simulation using a test like the Kolmogorov-Smirnov (K-S) test.  However, such tests are generally sensitive to any differences in the distributions and have a low level of discrimination in indicating precisely what that difference is.

What drew our attention to this result was the unexpected peak of the spin orientation for values of $\cos\theta < 0.2$.  Had we seen a peak in the range $0.2 < \cos\theta < 0.4$ we might have been equally impressed.  So the idea is to assess the observed distribution relative to a set of distributions that would have caught our attention.  We do this by introducing ``feature measures'' that measure the features that single out the histogram as being of particular interest, and ask how often we might have seen one that was at least as interesting in this respect.  The feature measure is perhaps more a measure of how justified we should feel in regarding the distribution as perceptually exceptional, rather than a measure of statistical significance.

% --------------------------------------------------------
\subsection{Feature measures}
% --------------------------------------------------------
We calculate, for the binned histogram of the distribution of spin-filament angles, three indices based on the $\cos\theta$ histogram (since it was the binned histogram that was the object that caught our appreciation).  The feature measures are
\begin{eqnarray}
 FM1   &=& bin(0.1) + bin(0.2)  \nonumber \\ 
			&& \qquad - 2 [bin(0.3) + bin(0.4)] +  \nonumber \\
			&& \qquad \qquad + bin(0.5) + bin(0.6)   
\label{eq:FM1}	\\		
 FM2   &=&  bin(0.8) +  bin(0.9) + bin(1.0) 
\label{eq:FM2} \\
 FM3   &=&  |FM1| + FM2 
\label{eq:FM3}
\end{eqnarray}
where $bin(n)$ refers to the occupancy of the bin $n-0.1 < \cos\theta < n$ with $0 < n < 1.0$.  The use of these particular measures in describing what we see is rather self-evident, but they can be formally derived from the data using component wavelet analysis of the histograms of the simulated distributions and the data.  

$FM1$ will be recognised as an estimator of the curvature of the interval $0 < \cos\theta < 0.6$.   $FM2$ is the occupancy of the last three bins.  $FM3$ is a composite index reflecting both these attributes.  Note that we have constructed $FM3$ using the modulus of $FM1$: we want $FM3$ to refelct any histigram that might have aroused our attention, not only the histogram we have derived from the data.  Using $|FM1|$ rather than $FM1$ in defining $FM3$ simply reduces the measured level of ``interestingness'' of our histogram.  Since these feature measures are in fact rather broad averages: they avoid sensitivity to details within the distribution and so they reflect broad features that would attract attention when viewing data.

The feature measure gives lower significance levels than would a simple test like the K-S test since it allows a greater number of accidental possibilities to be considered as being similar to or greater than deviations from the reference sample.

We shall make two comparisons.  The first with a sample using our SDSS Class A data in which the spin axis orientations have been randomly reassigned to the filaments in the same sample,  The second using a sample constructed from a model in which galaxies have randomly oriented spin axes relative to a sample of filaments having the same selection function as the observed SDSS sample.

% ---------------------------------------------------------
\subsubsection{A shuffled Class A sample}
% ---------------------------------------------------------
If we randomize the data by simply shuffling the spin axes in the sample among the filaments we get the distribution shown in figure \ref{fig:shuffletheta}.  The distribution of $\cos \theta$ is consistent with a uniform distribution.

This would lead us to believe that there is indeed a strong effect in the data.  For reference we also show the histogram that would arise if the spin axis orientation reltive to the host filament were random.  We have chosen a flat distribution for $\cos\eta$ with a cut-off at $\eta = 45^{0}$, as suggested by figure \ref{fig:lineofsight}.  

\begin{figure}
%  \mbox{\hskip -0.65truecm\includegraphics[bb=0 0 789 565,width=1.1\columnwidth]{figures_eps/FM1.png}} 
%  \mbox{\hskip -0.65truecm\includegraphics[bb=0 0 789 565,width=1.1\columnwidth]{jonesspin.fig12.png}} 
  \mbox{\hskip -0.65truecm\includegraphics[width=1.1\columnwidth]{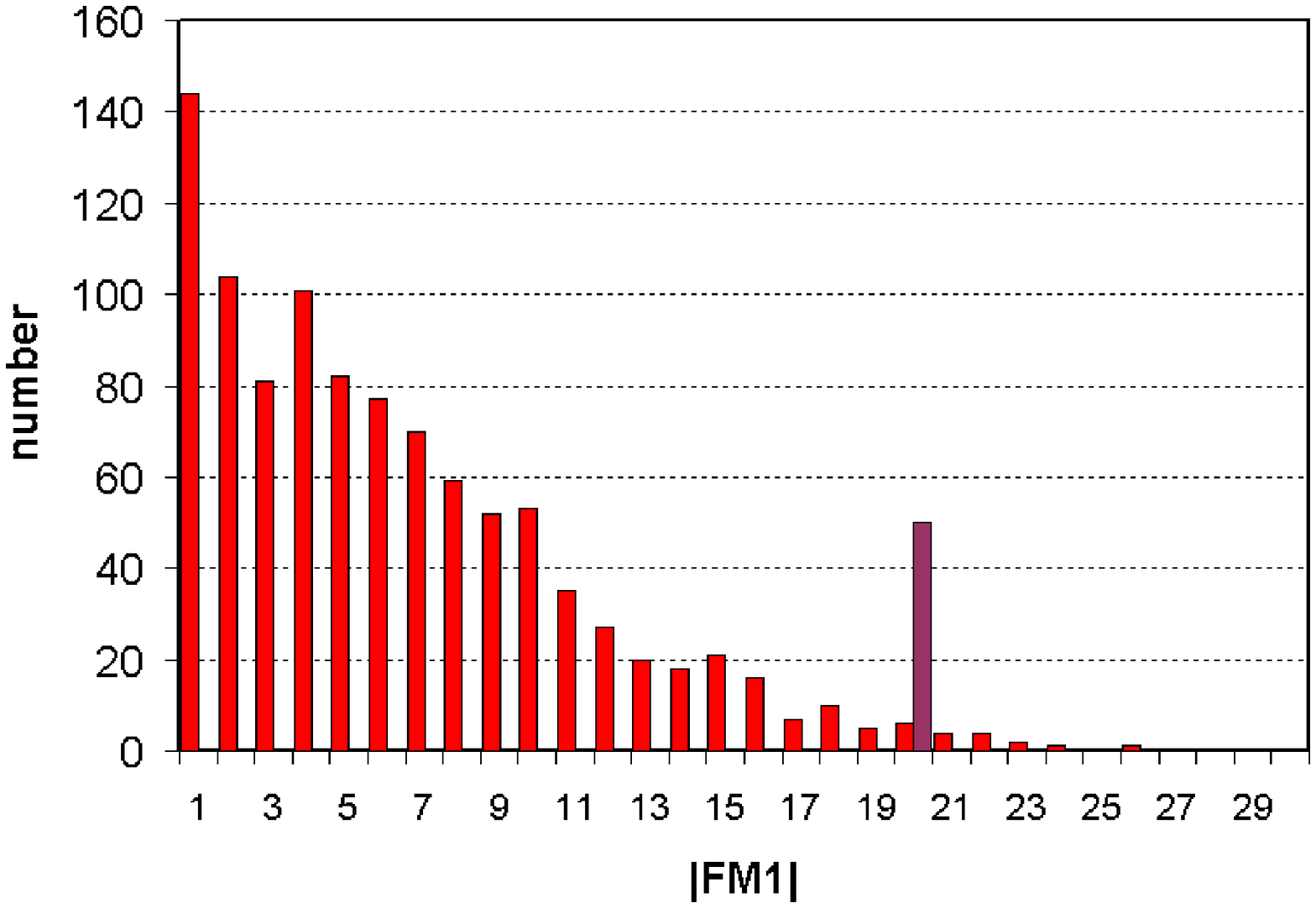}} 
  \caption{The absolute value of the curvature of the low $\cos\theta$ bins as indicated by our index $|FM1|$. Our SDSS sample is represented by the black spike, its height has no relevance except for clarity of display.  Our sample would be ranked joint 8th (along with 4 others) relative to the total sample of 1000 $\cos\theta$ values: the two-sided probability of getting a feature attracting our attention, like that in our analysis, by chance is about 1\%}
  \label{fig:FM1} 
%\end{figure} 
%\begin{figure}
%  \includegraphics[bb=0 0 791 566,width=\columnwidth]{figures_eps/FM1best.png} 
%  \includegraphics[bb=0 0 791 566,width=\columnwidth]{jonesspin.fig13.png} 
  \includegraphics[width=\columnwidth]{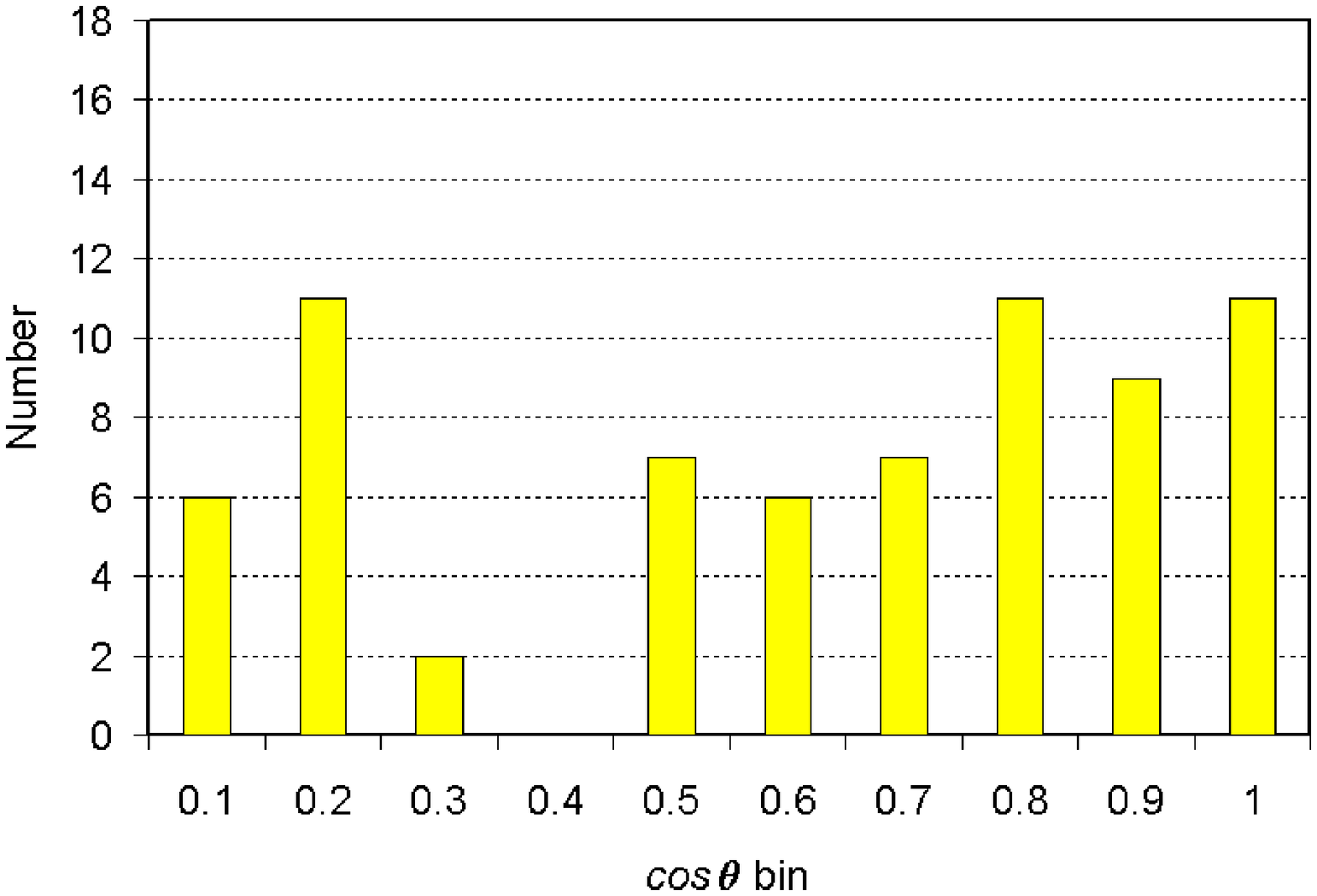} 
  \caption{The distribution in the reference sample having the highest value of feature measure $|FM1|$.}
  \label{fig:FM1best} 
%\end{figure} 
%\begin{figure}
%  \includegraphics[bb=0 0 789 541,width=\columnwidth]{figures_eps/FM1hump.png} 
%  \includegraphics[bb=0 0 789 541,width=\columnwidth]{jonesspin.fig14.png} 
  \includegraphics[width=\columnwidth]{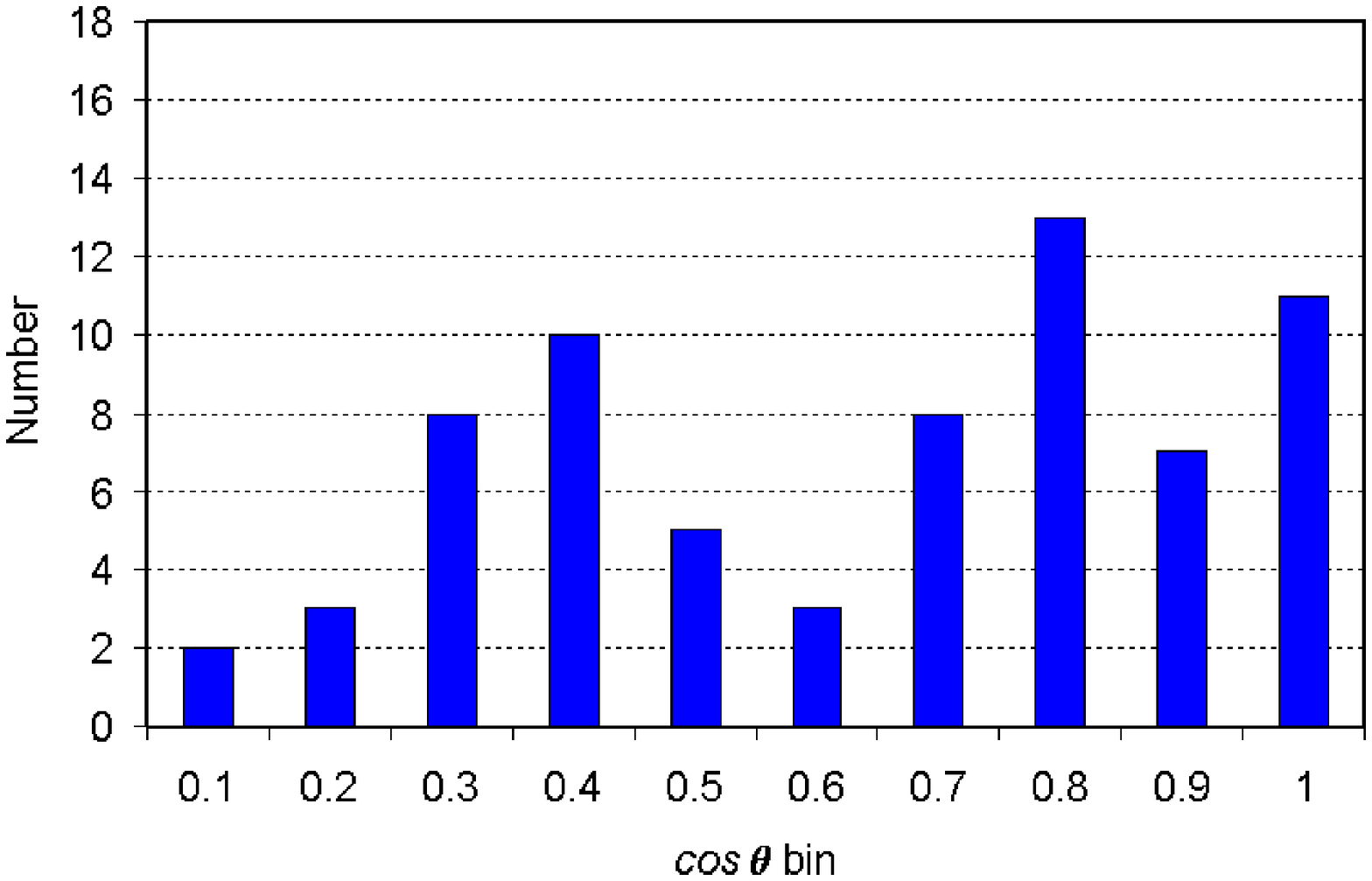} 
  \caption{A distribution in the reference sample having a feature that, according to the $|FM1|$ measure, would have been as noticeable as the feature shown in the data}
  \label{fig:FM1hump} 
\end{figure} 

% --------------------------------------------------------
\subsubsection{A simulated sample}
% --------------------------------------------------------
We simulated a random distribution of spin axis - filament angles, $\theta$, for a sample of filaments whose angle with the line of sight, $\eta$, followed a simple distribution like that indicated in figure \ref{fig:lineofsight} for the distribution of our Class A filaments.  We modelled this with a uniform distribution cut off at $\cos\eta > 45^{0}$.   

From this distribution we randomly generated 1000 samples, each containing 70 points with a value for $\theta$, and calculated the feature measures $FM1, FM2, FM3$ as per equations (\ref{eq:FM1}), (\ref{eq:FM2}), (\ref{eq:FM3}).  Since $FM2$ can be positive or negative, we used both $FM2$ and $|FM2|$ as measures.  The first of these is indicator of a valley in the low $\cos\theta$ distribution, while the second of these looks for any significant feature there, be it a valley or a hump.  The feature measures for the random sample were then ranked: the position of our sample in these rankings was taken as the measure of significance.

% --------------------------------------------------------
\subsubsection{First Remarks}
% --------------------------------------------------------
It will be noticed that the mean samples depicted in figures \ref{fig:shuffletheta} and \ref{fig:simulatetheta} are not quite the same: the distribution shown in figure \ref{fig:shuffletheta} is somewhat flatter than that shown in figure \ref{fig:simulatetheta}.  The former is based on a large number of random shuffles of the actual data while the latter is based simply on our model.  The difference almost certainly lies in the simplicity of the model selection function leading to figure \ref{fig:simulatetheta} where we have chose a sharp cut-off rather than a gradual one.

There is good reason for using two different comparison samples.  In the first sample, in which the Class A data is shuffled, there may be data selection issues that might themselves give rise to an anomalous distribution.  On the other hand, the artificially constructed sample, while random by construction, might not accurately reflect such anomalies.  The agreement between the two samples is encouraging and argues strongly that there is no anomaly arising from, say, the filament finding process.

It will also be noticed that in both figures it looks like there may be an excess of galaxies having spin angles with $\cos \theta > 0.8$.  That is one of several alternative explanations for the observed distributions.  We shall quantify this below using a feature measure designed to explore this feature.

% --------------------------------------------------------
\subsection{Results}
% --------------------------------------------------------
% --------------------------------------------------------
\subsubsection{$FM1$ and $|FM1|$}
% --------------------------------------------------------
Larger values of the feature measure $|FM2|$ indicate a significant hump or hole in the distribution of $\cos\theta$ relative to our reference sample.  The distribution of this statistic is shown in figure \ref{fig:FM1}, which shows that the chances of getting a feature as remarkable as the one in our sample is on the order of 1\%.  The mean value of $|FM1|$ is 6.37, whlie the value for our Class A sample is 21.  In particular, from $FM2$, the chance of getting a feature having the same shape as the data suggests is $< 0.5\%$.

We show two examples of high values of $|FM1|$.  In figure \ref{fig:FM1best} we show the distribution having the highest value of $FM1=23$.  The Class A sample has $FM1_{Class A} = 21$.  The other example, figure\ref{fig:FM1hump} shows an equally impressive feature having the opposite sign for $FM1$.  Such a distribution would have been considered remarkable and has to be taken into account when assessing how significant our result might be. 

\begin {table}
  \begin {center}
	Class A histogram ranking among samples of 1000 histograms
    \begin {tabular}{||l|c|c|c|c||}
      \hline
      Model &  $FM1$ & $|FM1|$ & $FM2$ & $FM3$ \\
      \hline
		Shuffled sample   & 7.5   &   16	& 1  & 1 \\
		Simulated sample  & 6.5   &   10	& 10 & 1 \\
      \hline
    \end {tabular}
    \caption{Rankings of Feature Measures of the Class A filament histograms sample relative to histograms from two randomised samples: a sample in which the actual data is shuffled and a sample that is constructed from a simple model.}
    \label{table:significance}
  \end {center}
\end {table}
% --------------------------------------------------------
\subsubsection{$FM2$}
% --------------------------------------------------------
The feature measure $FM2$ can reveal an excess of high $\cos\theta$ objects relative to the reference model. The SDSS sample of Class A filaments is ranked joint 8th on $FM2$ along with 6 others.  The expected value for $FM2$ is $\langle FM2 \rangle = 29$ for a sample of this size, while the value for our Class A filaments is $FM2_{Class A}=39$.  The high $\cos\theta $  part of the SDSS sample appears to be unusually high (probability $< 1\%$).  Of course that might be expected since if there is a hollow in the distribution relative to the model, there has to be a compensating excess somewhere else.  

We show in figure  \ref{fig:FM2examples} examples of distributions of distributions having equal or higher $FM2$-ranked values.  The flatness of the region where $\cos\theta > 0.7$ in these simulations is not surprising since the distribution would be flat for a uniform distribution of orientations.  Censoring the angle $\eta$ that the filament makes with the line of sight removes filaments in such a way as to approximately preserve the expected flatness of the $\cos\theta$ distribution for larger values of $\cos\theta$.  

We note from the sample shown in figure \ref{fig:FM2examples} having the highest value of of $FM1$, that the low $\cos\theta$ distribution is weaker than in our Class A sample.  It is because of that that the composite index $FM3$ for our Class A sample is considerably higher than for any sample in the reference distribution.

\begin{figure}
  \includegraphics[width=\columnwidth]{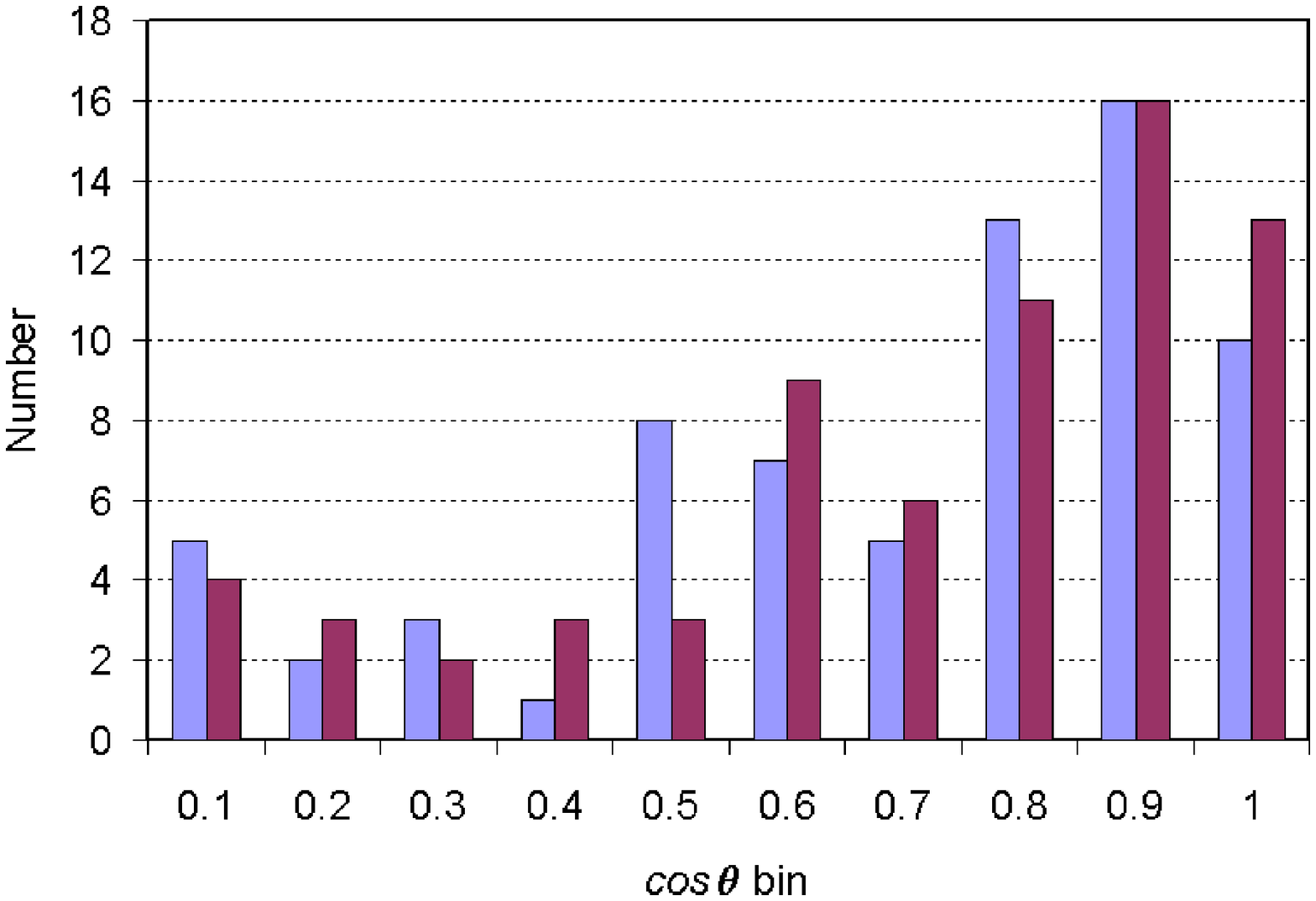} 
  \caption{Two examples from the reference distributions having equal or higher values of the feature measure $FM2$ than the Class A sample.  One sample was selected because it has the highest value of $FM1$ in that set, the other because it has the highest $FM2$ with a positive $FM1$}
  \label{fig:FM2examples} 
%\end{figure} 
%\begin{figure}
%  \includegraphics[bb=0 0 790 566,width=\columnwidth]{figures_eps/FM3best.png} 
%  \includegraphics[bb=0 0 790 566,width=\columnwidth]{jonesspin.fig16.png} 
  \includegraphics[width=\columnwidth]{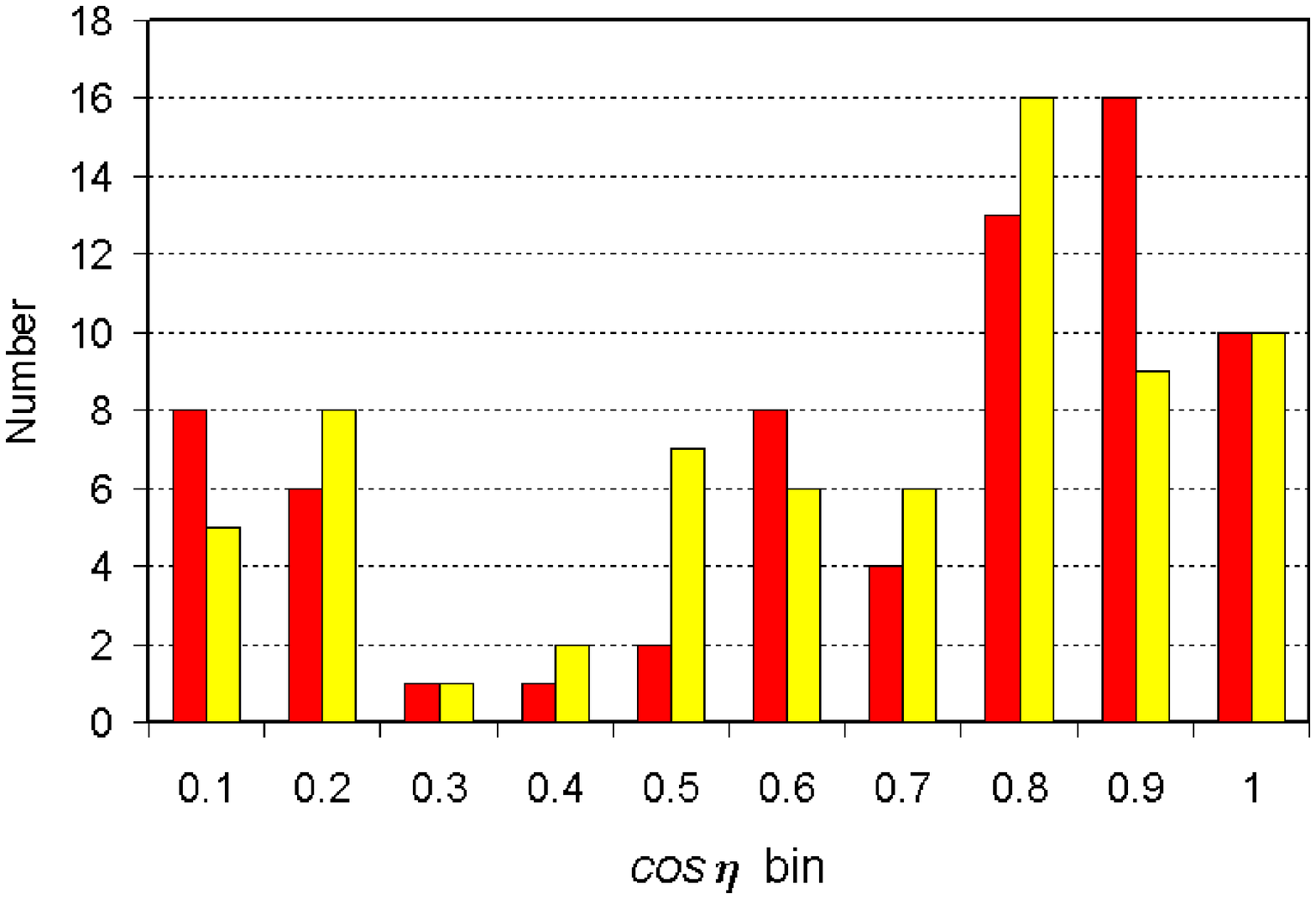} 
  \caption{The best of the reference models as gauged from the $FM3$ feature measure.  The similarity is remarkable: this attests to the effectiveness of using feature measures to assess this kind of data.}
  \label{fig:FM3best} 
\end{figure} 
% --------------------------------------------------------
\subsubsection{$FM3$}
% --------------------------------------------------------
$FM3$ is a composite measure assessing the entire distribution from the point of view of features.  The Class A SDSS sample ranks first on this measure by a long way: this is perhaps not surprising since the Class A sample is in the top $1\%$ of both $FM1$ and $FM2$.   However, $FM3$ allows us to select the ``best'' simulation from the reference sample, this is shown in figure \ref{fig:FM3best}.  The top $FM3$ samples include a variety of values of $FM1$ and $FM2$, so the selection of this particular example was taken from the highest ranking $FM3$ objects in such a way as to rank highly in both $FM1$ and $FM2$.  This is a rare example among the reference distributions and if, on that basis, we were to assess the chance of getting this at random it would be on the order of $0.1\%$.  However, any single instance of a general distribution is, by its very nature, rare, and we should not take that too seriously!

% --------------------------------------------------------
\subsection{Other statistical measures}
% --------------------------------------------------------
The Feature measures are somewhat \textit{ad hoc}, and that is why they are a measure on ``interestingness'' rather than a measure of ``significance''. We used a data mining approach to constantly pare down the sample to something useful without actually selecting for the very effect we were seeking to establish.  

% --------------------------------------------------------
\subsubsection{Significance tests}
% --------------------------------------------------------
The white histogram of figure \ref{fig:shuffletheta} shows the result of assigning the sample galaxies to a randomised sample of filaments.  The $\chi$-squared difference between this randomised histogram and the data (red) is 26.5 for 8 degrees of freedom.  The actual sample is very unlikely ($P < 0.1\%$) to have been drawn from a sample of galaxies that are randomly oriented relative to the parent filament.  There are many rearrangements of the histogram that would lead to as bad a chi-squared value: what the feature measure attempts to do is to single out the "interesting" rearrangements.

This situation arises because the $\chi$-square test takes no account of the distribution of the deviations: the underlying hypothesis is that the deviations are independently distributed random errors.  This is beautifully illustrated in Anscombe's incisive discussion in which he presents four quite different data samples distributions, \textit{Anscombe's Quartet}, having the same basic statistical properties \citep{Anscombe73}. 
%See also the \textit{Wikipedia} page for ``Anscombe's quartet''.  
The \textit{Feature Measure} is simply a way of carrying out Anscombe's admonishment for exploratory data analysis.

An alternative is to ask what is the probability that the Class A sample as shown in figure \ref{fig:dataangles} is drawn at random from the host sample of all 492 edge-on galaxies of figure \ref{fig:AllFilaments_theta}.  Applying the Mann-Whitney rank sum test to the distribution of $\cos\theta$ in the Class A sample and the total sample shows that the Class A sample is almost certainly not drawn randomly from the total sample (confidence level of the U-statistic $=6.9\sigma$).  

% --------------------------------------------------------
\subsection{Further remarks}
% --------------------------------------------------------
\subsubsection{Sample selection}
% --------------------------------------------------------
Most of the galaxies in the sample do not have $b/a < 0.2$ and that automatically gets rid of a large fraction of galaxies in the DR5 catalogue, leaving some 50,000.  However, most of these 50,000 or so edge-on galaxies do not lie on MMF filaments.  That reduces the number for this analysis to below 500: this is the sample of edge-on galaxies that lie in  MMF generated filaments.  We could have acquired a larger same by using a less restrictive axial ratio cut-off, like $b/a < 0.25$, but this might have introduced further uncertainties.

The classification that was done visually afterwards merely defined the subset of filaments that were considered, via a variety of criteria, to be reliable.  It turned out, with subsequent analysis, that this final selection favoured filaments that were both oriented some way from the line of sight and that were not at the extreme depths of the survey.

The question might arise as to the accuracy of the determinations of the various angles, especially their cosines.  This reflects on our choice of binning.   As figure \ref{fig:specials} shows, there should be little doubt about the accuracy of the value of the orientation of the galaxy relative to the sky.  The problem is in estimating the error in assessing the inclination of the filament to the line of sight.  There is no clear prescription for doing that, especially when the filament is not straight.  We ameliorated problems arising from estimation of filament orientation by including in our filament selection process the requirements that filaments  be straight and that galaxies should not be located at or near to filament branching points.  The slight relaxation of these requirements contributes to the difference between the Class A and Class B samples.

This type of selection is in fact a fundamental part of data mining: get rid of the objects that probably don't provide you with any information and look at what is likely to tell you something.  Again, that is why the focus of our attention is ``interestingness'' rather than ``significance''.  That should not make the analysis any less convincing, it merely emphasises the notion that there is something going on here and we should look at a bigger and better sample.

% --------------------------------------------------------
\subsubsection{Some additional comments}
% --------------------------------------------------------
Given the data, it is clear from this analysis that the distribution of spin orientations is anomalous.  All Feature measures are highly significant (better than the $1\%$ level).  Confirmation of that anomaly and establishment of the nature of the anomaly must await a bigger sample, but that does not stop us from speculating about what is going on!  However, when we come to samples like DR7, the volume of data will preclude visual assessment of the filaments and we will have to do this automatically.  That is perhaps the biggest challenge in using DR7.

% --------------------------------------------------------
\section{The nature of the  aligned objects}
\label{sec:objects}
% --------------------------------------------------------
What is clear is that there is a dearth of objects having spin orientations $0.2 < \cos\theta < 0.5$, and this manifests itself as either an excess of objects with $\cos\theta < 0.2$ or and excess of objects with $\cos\theta > 0.5$, or both.  This is a relatively narrow range of angles ($60^o < \theta < 80^o$), though it is difficult to make this range more precise with the data as it stands.

The 14 anomalous objects are tabulated in table \ref{table:candidates} and an instance from this table is shown in figure \ref{fig:specials}.  

The colour-magnitude plot for the entire sample of edge-on galaxies in the sample and for the candidate sample is shown in figure \ref{fig:colourmagnitude}.  
\begin{figure}
  \includegraphics[width=\columnwidth]{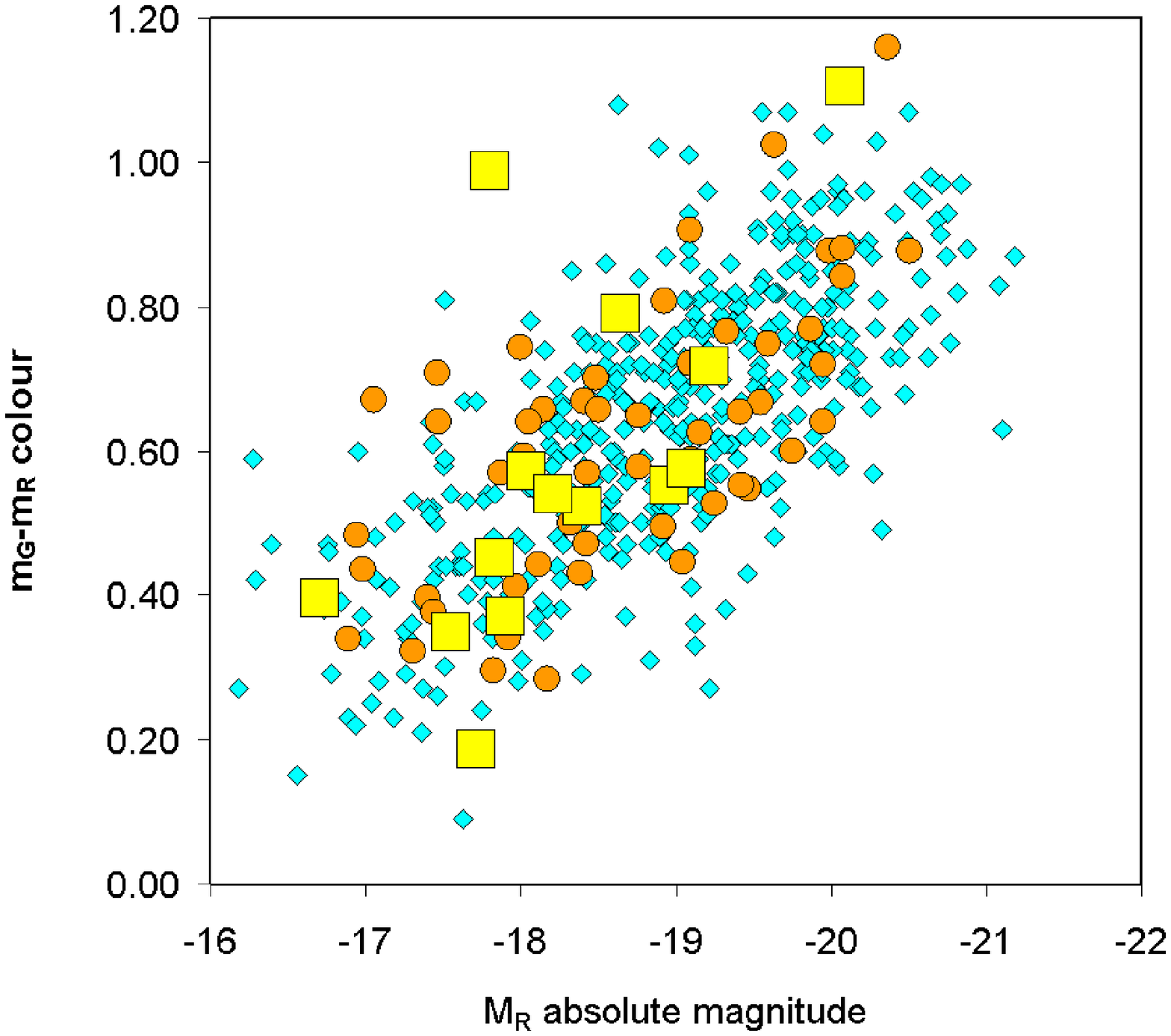} 
  \caption{Colour magnitude diagram for the entire sample of edge-on galaxies (dots) showing the gaalxies from the Class A filaments (circles) and the subsample of 14 candidate galaxies (squares).  The distributions are consistent with the hypothesis that the candidate sample they drawn randomly from the entire sample.}
  \label{fig:colourmagnitude} 
%\end{figure} 
%\begin{figure}
%  \includegraphics[bb=0 0 915 673,width=\columnwidth]{figures_eps/Diameter-Magnitude.png} 
%  \includegraphics[bb=0 0 915 673,width=\columnwidth]{jonesspin.fig18.png} 
  \includegraphics[width=\columnwidth]{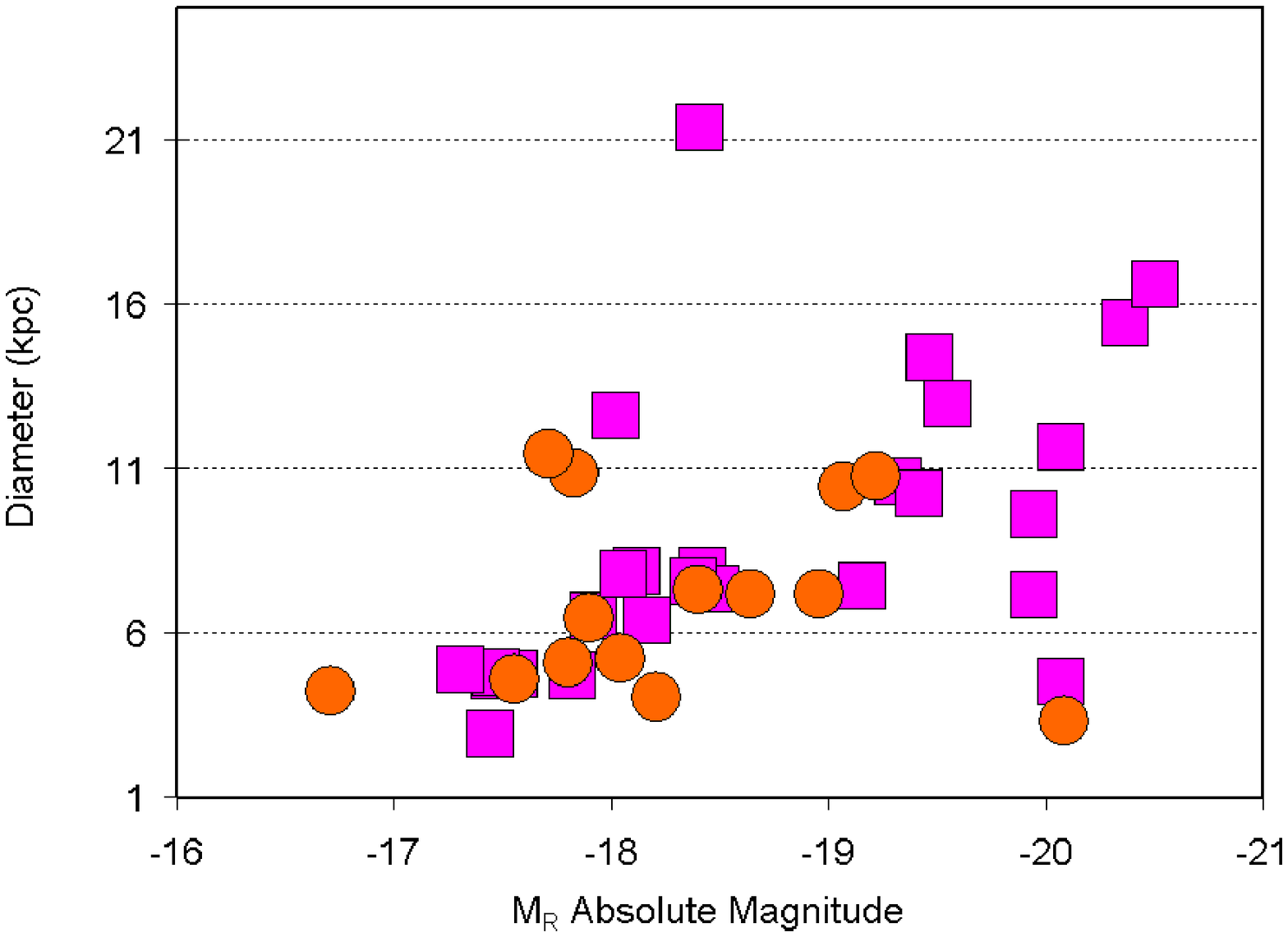} 
  \caption{The diameter - magnitude relationship for two sub-samples of edge-on galaxies taken from Class A filaments.  The red dots are the sample of 14 objects having spin axis - filament angle $\theta$ such that $\cos\theta < 0.2$, the blue boxes are a sample of galaxies having $\cos\theta > 0.8$.}
  \label{fig:diametermagnitude} 
\end{figure} 
There is no evidence on the basis of colours or spectra  that these galaxies are in any way special or peculiar.  It should be noted that we have not made any attempt to separate out early and late type galaxies.

The diameter - magnitude relationship is shown in figure \ref{fig:diametermagnitude} for two sub-samples of galaxies: one whose spin axes are perpendicular to their host filament, $\cos\theta < 0.2$, and another whose spin axes lie along the filament, $\cos\theta > 0.8$ .  Although the evidence is not strong, there is a suggestion that there is a lack of the bigger and the intrinsically brighter objects among the 14 ``specials'' we have identified.  If this were true it would be an important clue in coming to an understanding of this phenomenon.

\begin {table}
  \begin {center}
    \begin {tabular}{||l|c|c|c|c|c|c||}
      \hline
      ID &  $\cos \theta$ & RA & dec & $z$ & $M_g$ & $M_r$ \\
      \hline
		198	    & 0.0024   &   244.108	& 25.6138 & 0.0402 & -17.83 & -17.83 \\
		8$^a$   & 0.0052   &   132.965	& 40.8162 & 0.0293 & -17.55 & -17.55 \\
		289	    & 0.0201   &   138.069	& 51.6174 & 0.0281 & -18.95 & -18.95 \\
		150$^b$	& 0.0461   &   160.263	& 39.9217 & 0.0685 & -20.09 & -20.09 \\
		345	    & 0.0569   &   182.351	& 45.5697 & 0.0669 & -19.07 & -19.07 \\
		242	    & 0.0600   &   188.740	& 39.9192 & 0.0569 & -18.40 & -18.40 \\
		95	    & 0.0683   &   213.151	& 51.3609 & 0.0750 & -19.22 & -19.22 \\
		317	    & 0.0777   &   142.991	& 33.2113 & 0.0423 & -17.71 & -17.71 \\
		50$^c$	& 0.1020   &   145.122	& 41.4512 & 0.0471 & -18.04 & -18.04 \\
		193	    & 0.1419   &   242.538	& 26.9232 & 0.0319 & -18.64 & -18.64 \\
		89	    & 0.1472   &   224.607	& 48.7573 & 0.0311 & -17.90 & -17.90 \\
		424	    & 0.1596   &   121.679	& 47.3122 & 0.0226 & -16.71 & -16.71 \\
		84	    & 0.1674   &   190.719	& 55.1458 & 0.0160 & -18.21 & -18.21 \\
		348	    & 0.1696   &   216.877	& 40.9637 & 0.0184 & -17.80 & -17.80 \\
%		101$^d$	& 0.1719   &   212.948	& 52.7963 & 0.0747 & -19.68 & -19.68 \\
      \hline
    \end {tabular}
    \caption{\small{The 14 candidate galaxies that provide our evidence for alignment processes in filaments.  They are listed in order the angle the spin axis makes with the host filament.  $M_r$ and $M_g$ denote the absolute magnitudes of the object in the SDSS $r$ and $g$ bands, calculated using the redshift $z$. The ID is the filament number in the FILCAT-0 catalogue of filaments found using the MMF technique \citep{Aragon07_thesis}.} All of these objects, with the exception of galaxy 101,  look like normal edge-on galaxies. However the following should be noted:  $^a$tidally disturbed,  $^b$the SDSS published diameter looks wrong so this datum is uncertain,  $^c$has a close-by neighbour of unknown redshift, $^d$not an edge-on galaxy.}
    \label{table:candidates}
  \end {center}
\end {table}
% =========================================================
\section{Conclusions and Discussion}
\label{sec:conclusions}
% ---------------------------------------------------------
We have taken a substantially different approach to the problem of finding evidence for non-randomness in the orientation of galaxies.  The process we have used has been almost archaeological or forensic in nature in that we look for likely spots for evidence of tidal influence and then search for objects of a specific kind.

Firstly we have chosen to look at the relationship between the spin axis of the galaxies and the filamentary structures that host them.  In so doing we are looking where we expect to find evidence.  A clustered environment seems less promising since galaxy interactions and orbital dynamics will have played a role, whereas we would not expect to see appreciable local dynamical influences in filaments. Moreover, the filaments are relatively easy to find using techniques like MMF based on DTFE density sampling.  Indeed, we would argue that using a less effective filament finder might hinder the discovery any such alignment effects.

\begin{figure*}
  \includegraphics[width=\textwidth]{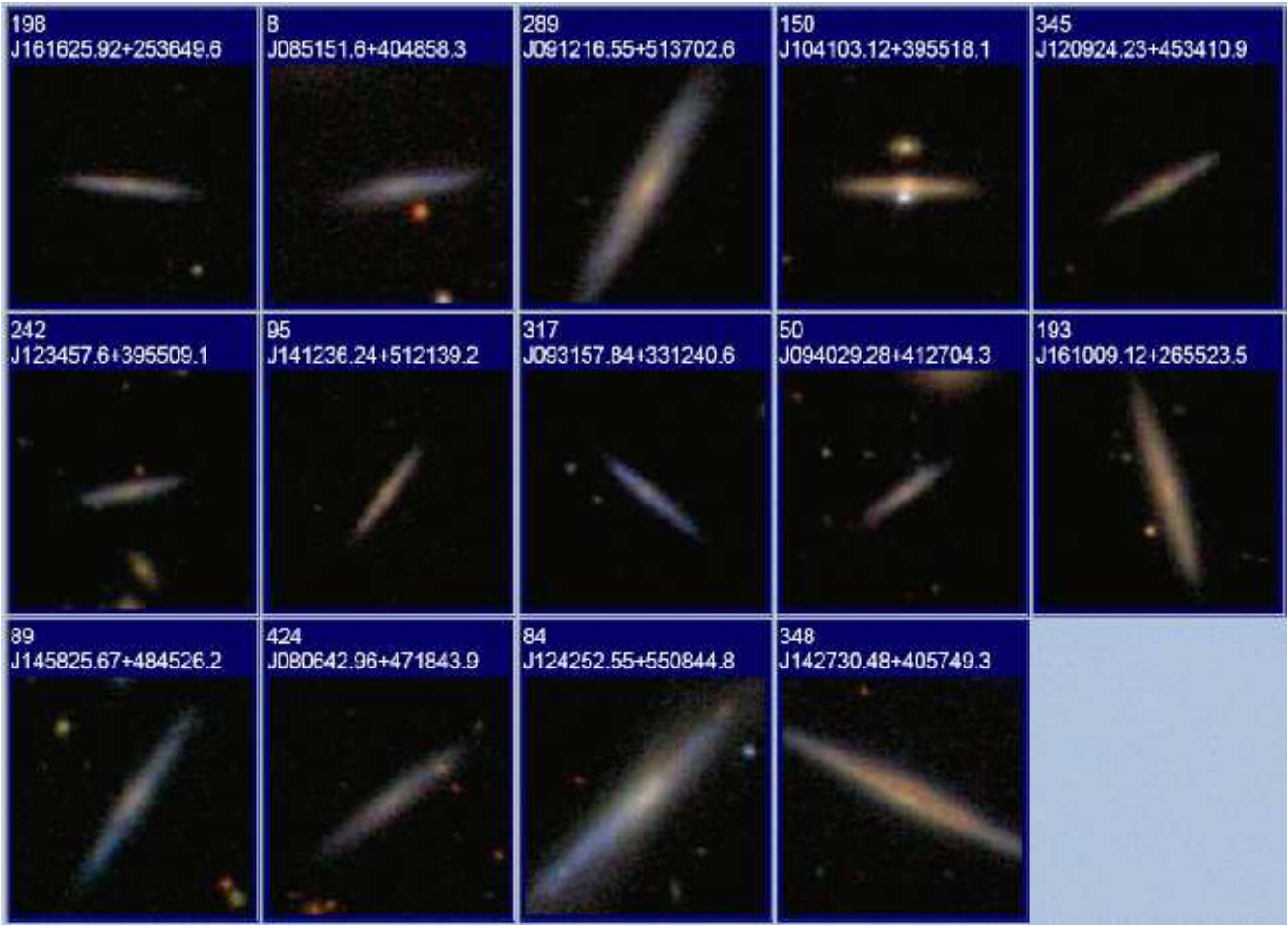} 
  \caption{The 14 aligned filament galaxies (see table 2). Images from the SDSS database.}
  \label{fig:specials} 
\end{figure*} 

The filament data was pre-processed so as to correct for the finger-of-god effects, and then thinned so as to bring the member galaxies closer to the spine of the filament.  Then the catalogue of filaments was censored so as to give a smaller, better defined sample of filaments, but without regard to the orientation of the galaxies lying in the filaments.  In making this selection we studied the finger of god corrected maps in which the distribution had been thinned, and subjectively eliminated objects that were complex, having branching structures.  This reduced the original sample of 426 filaments to 67 ``Class A'' filaments.  Most contained but one edge-on galaxy: only three pairs of our sample of 69 galaxies in Class A filaments lie in the same filament and no filament contains more than one of our ``special'' subgroup of 14 galaxies.  No Class A filament contained more than two edge on galaxies.  

We suggest that the distribution is not consistent with a random distribution of spin orientations and, in particular, we find 14 objects among these 67 filaments that are oriented perpendicular to their host filament.  There is corroborating evidence for this assertion,  The slightly wider sample of 121 Class A and Class B objects gave some support for this, as does the sub-sample of filaments lying in the plane of the sky.

In order to assess the statistical standing of this conclusion we characterise the distribution of spin angles by a number of ``feature measures'' and assess the chances of coming up with values for those measures equal to or greater than the values for our Class A SDSS sample.  Both shuffling the spins in the Class A sample among their filaments and using a sample of 1000 histograms for randomly distributed orientations shows that our observed Class A distribution is indeed remarkable.

If we take this analysis at face value, there are at least two possible interpretations of the histogram we find for the spin orientations.  Either there is an excess of objects having their spin axes perpendicular to the spine of the host filament, or there is an excess of filaments having their spin axes parallel to the spine.  These views are not inconsistent: it could simply be that the distribution of orientations is bimodal.  It is easier to follow up the first of these alternatives since there are only 14 objects involved, while in the second of the alternatives there is no way of deciding which galaxies might individually be responsible for the excess.

The distribution of the relative orientations of the spin axis of the galaxy and its host filament provided us with a sample of 14 candidate galaxies that we would claim provide evidence for such tidal interactions.  These candidates are typical of any other edge-on galaxy in the entire sample - they are not in any way distinguished except by their unexpected orientation and perhaps in being slightly smaller than other edge-on galaxies in the sample.

It turns out that these candidate galaxies lie centrally in relatively insignificant filaments.  This may suggest that they have suffered less from interactions with other galaxies and have been influenced mostly by the global tidal field exerted by their parent filament.

It might occasion surprise that an original sample of over 50,000 galaxies has been whittled down to a mere 14!  The first sample of filaments was defined using the entire galaxy sample, and we selected only those containing edge-on galaxies defined as having axial ratio $b/a < 0.2$.  This left us with only 426 filaments, which subsequent quality assessment reduced to 67 containing 69 edge-on galaxies.

Obviously this should now be repeated using a sample like the DR7 release of the SDSS.  If these 14 galaxies are indeed fossils of earlier tidal interactions, using a bigger catalogue would either give many more candidate objects or eliminate the effect entirely.  One advantage of a bigger sample is that we could make a flatter axial ratio cut.  The general procedure is, however, far less straightforward since it would not be possible to do the visual triage of filaments to select out ``clean'', Class A, filaments.  This would have to be done algorithmically.

However, until that is done, we offer this study as a set of fossils providing forensic evidence that tidal interactions have affected the orientations of the spin axes of galaxies.  

% =========================================================
\section*{Acknowledgements}
BJ gratefully acknowledges the hospitality of the Kapteyn Astronomical Institute in Groningen, 
and to his collaborators for their remarkable patience in getting various parts of this paper completed. 
MA thanks Alex Szalay for invaluable help towards exploring the SDSS database in the early stages of 
this project. BJ and RvdW are particularly indebted to Niels Bos for his invaluable help in producing 
figure D1. We are also grateful to J~.M.~van der Hulst and P.~Sackett for encouraging and useful 
discussions during the defining stages of this project. We thank the anonymous referee for the 
careful and helpful appraisal of our manuscript.  

Funding for the SDSS and SDSS-II has been provided by the Alfred P.
   Sloan Foundation, the Participating Institutions, the National
   Science Foundation, the U.S. Department of Energy, the National
   Aeronautics and Space Administration, the Japanese Monbukagakusho,
   the Max Planck Society, and the Higher Education Funding Council for
   England. The SDSS Web Site is http://www.sdss.org/. The SDSS is managed 
   by the Astrophysical Research Consortium for the
   Participating Institutions. The Participating Institutions are the
   American Museum of Natural History, Astrophysical Institute Potsdam,
   University of Basel, University of Cambridge, Case Western Reserve
   University, University of Chicago, Drexel University, Fermilab, the
   Institute for Advanced Study, the Japan Participation Group, Johns
   Hopkins University, the Joint Institute for Nuclear Astrophysics,
   the Kavli Institute for Particle Astrophysics and Cosmology, the
   Korean Scientist Group, the Chinese Academy of Sciences (LAMOST),
   Los Alamos National Laboratory, the Max-Planck-Institute for
   Astronomy (MPIA), the Max-Planck-Institute for Astrophysics (MPA),
   New Mexico State University, Ohio State University, University of
   Pittsburgh, University of Portsmouth, Princeton University, the
   United States Naval Observatory, and the University of Washington." 

% ---------------------------------------------------------

\appendix
\section{The tidal shear field}
\label{app:tides}
%.............................
The relation between the fluctuating component of the density field, $\delta({\bf r},t)$, and gravitational potential $\Phi$ is established through the Poisson-Newton equation:
\begin{eqnarray}
& \nabla^2 \Phi = 4 \pi G \bar{\rho}_{m}(t) a(t)^2 \ \delta({\bf r},t) , \\
& \delta({\bf r},t)= \displaystyle\frac{\rho({\bf r})-\rho_{\rm u}}{\rho_{\rm u}}.
\end{eqnarray}
Here $\bar{\rho}_{m}(t)$ is the mean density of the mass in the universe that can cluster (dark matter and baryons).  

The peculiar gravitational acceleration due to the integrated effect of all matter fluctuations in the Universe is related to $\Phi({\bf r},t)$ through ${\bf g}=-\nabla \Phi/a$: 
\begin{equation}
{\bf g}({\bf r},t)\,=\,- 4\pi G \bar{\rho}_m
(t)a(t) \int {\rm d}{\bf r}'\,\delta({\bf r}^\prime,t)\,{\displaystyle
({\bf r}-{\bf r}^\prime) \over \displaystyle |{\bf r}-{\bf
r}^\prime|^3}\ .
\label{eq:gravstab} 
\end{equation}
and drives the peculiar (non-Hubble expansion components) of the cosmic motion.  The cosmological density parameter $\Omega_m (t)$ is defined by $\rho_{\rm u}$, via the relation $\Omega_m H^2 = (8\pi G/3) \bar{\rho_{m}} $ in terms of the Hubble parameter $H$. 

The tidal shear arising from this acceleration is given by the (traceless) tidal tensor $T_{ij}$ \citep{Weyedb1996},
\begin{eqnarray}
T_{\rm ij}&\,\equiv\,&-{\displaystyle 1 \over \displaystyle 2 a}\, 
\left\{ {\displaystyle \partial g_{{\rm b},i} \over \displaystyle \partial x_i} +
{\displaystyle \partial g_{{\rm b},j} \over \displaystyle \partial x_j}\right\}\,+\,
{\displaystyle 1 \over \displaystyle 3 a} (\nabla \cdot {\bf g}_{\rm b}) 
\,\delta_{\rm ij}
\end{eqnarray}
This can be expressed directly in terms of the fluctuation component of the density field, $\delta$, via the equation
\begin{eqnarray}
T_{ij}({\bf r})={\displaystyle 3 \Omega H^2 \over \displaystyle 8\pi}
\int \delta({\bf r}') {\cal{Q}}_{ij} {\rm d}{\bf r} - {\frac{1}{2}}\Omega H^2\ \delta({\bf r},t)\ \delta_{ij} .  \\
{\cal{Q}}_{ij} = \left\{{\displaystyle 3 (r_i'-r_i)(r_j'-r_j) -
|{\bf r}'-{\bf r}|^2\ \delta_{ij} \over \displaystyle |{\bf r}'-{\bf r}|^5}\right\}
\label{eq:quadtide}
\end{eqnarray}
This expression shows explicitly that source of the tidal field is the quadrupole component of the fluctuating matter distribution, ${\cal{Q}}_{ij}\,\delta({\bf r}') $.  The quadrupole component of the field falls of as $r^{-3}$, so, unlike the isotropic component, it is localised.  However, there is a considerable contribution from the largest scales as can be seen in the study of void alignments by \citet{Platen08} and in the study of shear fields in relation to gravitational lensing \citep{Gunn67, Wittam00}.  This can produce systemic shear that is correlated over large scales, resulting in systemic 
alignments that survive until mergers and tidal forces from other nearby structures start playing a role.

The tidal shear tensor has been the source of intense study by the gravitational lensing community since it is now possible to map the distribution of large scale cosmic shear using weak lensing data. Examples are the studies by \citet{HirSel04} and \citet{massey2007}.

\section{Tidal Alignment of Galaxy Spins}
\label{app:tidalign}
One may obtain insight into the issue of the alignment of a halo's angular momentum ${\bf L}$ with the surrounding 
matter distribution by evaluating the expression for the angular momentum ${\bf L}$,  
\begin{equation}
L_i \propto \epsilon_{ijk} T_{jm} I_{mk},
\label{eq:angmom}
\end{equation}
(with $\epsilon_{ijk}$ the Levi-Civita symbol) in the principal axis frame of the tidal tensor $T_{ij}$, 
\begin{eqnarray}
L_1&\,\propto\,&(\lambda_2-\lambda_3)\,I_{23}\nonumber\\
L_2&\,\propto\,&(\lambda_3-\lambda_1)\,I_{31}\\ 
L_3&\,\propto\,&(\lambda_1-\lambda_2)\,I_{12}\nonumber\,,
\end{eqnarray}
where $(\lambda_1,\lambda_2,\Lambda_3)$ are the tidal field eigenvalues. Since, by definition, $(\lambda_3-\lambda_1)$ is the 
largest coefficient, in a statistical sample of halos $L_2$ will on average get the largest contribution if we assume that 
the tidal and inertial tensors $T_{ij}$ and $I_{ij}$ are uncorrelated. 

However, the assumption above is not well justified, and in fact the two tensors are found to be correlated \citep{Porciani02}. 
Recently, \citet{Lee09} numerically determined the two-point correlations of the three eigenvalues of the nonlinear traceless 
tidal field in the frame of the principal axes of the tidal field. The numerical findings indicate that the correlation functions of 
the traceless tidal field and the density field are all anisotropic relative to the principal axes. Interesting is also 
that their correlations have much larger correlation length scales than that of the density field and increase along the directions 
normal to the first principal axes of the tidal field.

To incorporate the effect of a correlation between the tidal and inertial tensors $T_{ij}$ and $I_{kl}$, 
\cite{LeePen00} and \cite{LeePen01} suggested an analytical formulation involving a useful parametrization of 
the correlation. This they accomplished by writing down an equation for the autocorrelation tensor of the 
angular momentum vector in a given tidal field, averaging over all orientations and magnitudes of the inertia tensor. 
Arguing that the isotropy of the underlying density distribution allows the replacement of the statistical quantity 
$\langle I_{mq} I_{ns}\rangle$ by a sum of Kronecker deltas, one is left with the result that the autocorrelation tensor of 
the angular momentum vector is given by  
\begin{equation}
\langle L_i L_j | {\bf T} \rangle \propto {1 \over 3} \delta_{ij} + ({1 \over 3} \delta_{ij} - T_{ik}T_{kj} )
\label{eq:indepIT}
\end{equation}
If it is asserted that the moment of inertia and tidal shear tensors were uncorrelated, we would have only the 
first term on the right hand side, ${1 \over 3} \delta_{ij}$: the angular momentum vector would be isotropically 
distributed relative to the tidal tensor and one would not expect the presence of any significant 
galaxy spin alignments. Recognizing that the inertia and tidal tensors in general are not mutually independent, 
\cite{LeePen00,LeePen01} introduced a parameter $c$ to quantify this correlation, 
\begin{equation}
\langle L_i L_j | {\bf T} \rangle \propto {1 \over 3} \delta_{ij} + c ({1 \over 3} \delta_{ij} - T_{ik}T_{kj} )
\end{equation}
where $c = 0$ for randomly distributed angular momentum vectors.  The case of mutually dependent tidal and inertia tensors is 
described by $c = 1$ (see equation \ref{eq:indepIT}). Finally, they introduce a different parameter $a = 3c/5$ and write
\begin{equation}
\langle L_i L_j | {\bf T} \rangle \propto {{1+a} \over 3} \delta_{ij} - a T_{ik}T_{kj} 
\end{equation}
which forms the basis of much current research in this field. The value derived from a recent study of the Millennium simulations 
by \citet{LeePen07} is $a \approx 0.1$. 

\cite{LeeErdog07} quantified the preferential alignment of the angular momentum vector ${\bf L}$ along the intermediate 

principal axis ${\bf T_2}$ of the tidal tensor, by deriving the corresponding probability distribution $p(\theta_2)$ for 
the alignment angle $\cos{\theta_2}=|{\bf L}\cdot{\bf T_2}|/|{\bf L}|$. Including the effect of the correlation between 
the tidal and inertial tensor by means of the parameter $c$, the find that 
\begin{equation}
p(\cos{\theta_2})\,=\,{\displaystyle (1+c)\,\sqrt{1-{\displaystyle c \over \displaystyle 2}} \over \displaystyle 
\left[1\,+\,c\left(1-{3 \over 2}\cos^2{\theta_2}\right)\right]^{3/2}}\,,
\end{equation}
where $\cos{\theta_2}$ is assumed to be in the range $[0,1]$. If $c=0$, the probability distribution 
will be a uniform distribution $p(\theta_2)=1$. 

% --------------------------------------------------------------
\section{Finding filaments with MMF}
\label{app:mmf}
% --------------------------------------------------------------
One of the key issues in this investigation is to rigorously identify filaments in the large scale structure.  The Multiscale Morphology Filter, MMF, that we use to identify filamentary structures is presented in detail in \citep{Aragon07_MMF} where a detailed step-by-step description of the MMF algorithm can be found.  In this Appendix we briefly summarize the steps involved in the morphological segmentation of the cosmic web obtained from the N-body cosmological simulation.  

% --------------------------------------------------------------
\subsection{Scale Space}
% --------------------------------------------------------------
The DTFE density field $f_{\tiny{\textrm{DTFE}}}$ is the starting point of the morphological segmentation. 
The density field is smoothed over a range of scales by means of a hierarchy of spherically symmetric Gaussian 
filters $W_{\rm G}$ having different widths $R_n$. The $n^{th}$ level smoothed version of the DTFE reconstructed field 
$f_{\tiny{\textrm{DTFE}}}$ is assigned $f_n$, 
\begin{equation}
f_{\rm n}({\vec x}) =\, \int\,{\rm d}{\vec y}\,f_{\tiny{\textrm{DTFE}}}({\vec y})\,W_{\rm G}({\vec y},{\vec x})\nonumber
\end{equation}
where $W_{\rm G}$ denotes a Gaussian filter of width $R_n$: 
\begin{equation}
W_{\rm G}({\vec y},{\vec x})\, = \,{1 \over ({2 \pi} R^2)^{3/2}}\, \exp \left(- {|{\vec y}-{\vec x}|^2 \over 2 R_n^2}\right)\,.
\label{eq:filter}
\end{equation}

Scale Space itself is constructed by stacking these variously smoothed data sets, yielding the family $\Phi$ of smoothed density maps $f_n$:
\begin{equation}
\label{eq:scalespace}
\Phi\,=\,\bigcup_{levels \; n} f_n 
\end{equation}
A data point can be viewed at any of the scales where scaled data has been generated.  The crux of the concept of Scale Space is that the 
neighbourhood of a given point will look different at each scale.  There are potentially many ways of making a comparison of the scale 
dependence of local environment. We address the local ``shape'' of the density field. 

% --------------------------------------------------------------
\subsection{Local Shape}
% --------------------------------------------------------------
The local shape of the density field at any of the scales $R_n$ in the Scale Space representation of the density field can be quantified on the basis of the Hessian matrix, ${\tilde {\mathcal H}}_{ij}=\nabla_{ij} f_n({\bf x})$, 
\begin{eqnarray}
\frac{\partial^2}{\partial x_i \partial x_j} f_n({\vec x})\,=\,f_{\tiny{\textrm{DTFE}}}\,\otimes\,\frac{\partial^2}{\partial x_i \partial x_j} W_{\rm G}(R_{\rm n})\nonumber & \\
= \int\,{\rm d}{\vec y}\,f({\vec y})\,\,\frac{(x_i-y_i)(x_j-y_j)-  \delta_{ij}R_{\rm S}^2}{R_{\rm S}^4}& W_{\rm G}({\vec y},{\vec x}) \quad
\end{eqnarray} 
where $\{x_1,x_2,x_3\}=\{x,y,z\}$ and $\delta_{ij}$ is the Kronecker delta. In other words, at each level $n$ of the scale space representation the Hessian matrix is evaluated by means of a convolution with the second derivatives of the Gaussian filter, also known as the Marr (or, less appropriately, ``Mexican Hat'') Wavelet. In order to properly compare the values of the Hessian arising from the differently scaled variants of the data that make up the Scale Space, the Hessian is renormalized, $\tilde {\mathcal{H}}\,=\,R_{\rm S}^2 \,\mathcal{H}$, where $R_s$ is the filter width that has been used. 

The eigenvalues $\lambda_i$ of the Hessian matrix determine the local morphological signal, dictated by the local shape of the 
density distribution. A small eigenvalue indicates a low rate of change of the field values in the corresponding eigen-direction, and vice versa.
We denote these eigenvalues by $\lambda_{a}(\vec{x})$ and arrange them so that $ \lambda_1 \ge \lambda_2 \ge \lambda_3 $:
\begin{eqnarray}
\qquad \bigg\vert \; \frac{\partial^2 f_n({\vec x})}{\partial x_i \partial x_j}  - \lambda_a({\vec x})\; \delta_{ij} \; \bigg\vert  =& 0,  \quad a = 1,2,3 \\
\mathrm{with} & \quad \lambda_1 > \lambda_2  >  \lambda_3 \nonumber
\end{eqnarray}
The $\lambda_{i}(\vec{x})$ are coordinate independent descriptors of the behaviour of the density field in the locality of the point $\vec{x}$ and can be combined to create a variety of morphological indicators. The criteria we used for identifying a local bloblike, filamentary or sheetlike morphology are listed in table~\ref{tab:chap3_eigen_ratios}. 

% ======= TABLE 1 =========
\begin{table}
\label{tab:chap3_eigen_ratios}
\centering
%\begin{large}
\begin{tabular} {|l|c|l|}
\hline
\hline
%&&\\
Structure & $\lambda$ ratios & $\quad$ $\lambda$ constraints \\
%&&\\
\hline
%&&\\
Cluster &  $\lambda_1 \simeq \lambda_2 \simeq \lambda_3$ & $\lambda_3 <0\,\,;\,\, \lambda_2 <0 \,\,;\,\, \lambda_1 <0 $  \\
Filament     &  $\lambda_1 \simeq \lambda_2 \gg    \lambda_3$ & $\lambda_3 <0 \,\,;\,\, \lambda_2 <0  $  \\
Sheet    &  $\lambda_1 \gg    \lambda_2 \simeq \lambda_3$ & $\lambda_3 <0 $    \\
%&&\\
%\hline
\hline
\end{tabular}
%\end{large}
\vskip 0.25truecm
\caption{Behavior of the eigenvalues for the characteristic morphologies. The
         lambda conditions describe objects with intensity higher that their
	 background (as clusters, filaments and walls).  From the constraints 
         imposed by the $\lambda$ conditions we can describe the blob morphology  
         as a subset of the line which is itself a subset of the wall.}
\end{table}

% --------------------------------------------------------------
\subsection{Multiscale Structure Identification}
% --------------------------------------------------------------
In practice, we are interested in the local morphology as a function of scale. In order to establish how it changes with scale, we evaluate the eigenvalues and eigenvectors of the renormalised Hessian $\tilde {\mathcal{H}}$ of each dataset in the Scale Space $\Phi$. 

Since we are looking for three distinct structural morphologies - blobs, walls and filaments - the practical 
implementation of the segmentation consists of a sequence of three stages. Because curvature components are 
used as structural indicators, the blobs need to be eliminated before looking for filaments, after which the filaments 
have to be eliminated before looking for walls. This results in the MMF procedure following the sequence 
``clusters $\rightarrow$ filaments $\rightarrow$ walls''. At each of these three steps, the regions and scales are 
identified at which the local matter distribution follows the corresponding eigenvalue signature. 

In practice, the MMF defines a set of morphology masks, morphology response filters and morphology filters for each 
of the three different morphological components: clusters, filaments and walls. Their form is dictated by the particular 
morphological feature they seek  to extract, via the eigenvalues at each level in scale space and the criteria for each of the corresponding morphologies (table~\ref{tab:chap3_eigen_ratios}). The local value of the the morphology response depends on the local shape and spatial coherence of the density field. The morphology signal at each location is then defined to be the one with the maximum response across the full range of smoothing scales. 

The end result is a map segmented in clusters, filaments and walls that have been identified as most outstanding, and which vary in scale over the full range of scales represented in the Scale Space. 

% --------------------------------------------------------------
\section{Distribution of random spins with censored filament orientations}
\label{app:randomspins}
% --------------------------------------------------------------
\begin{figure*}
   \vskip -0.25truecm
  \includegraphics[width=0.95\textwidth]{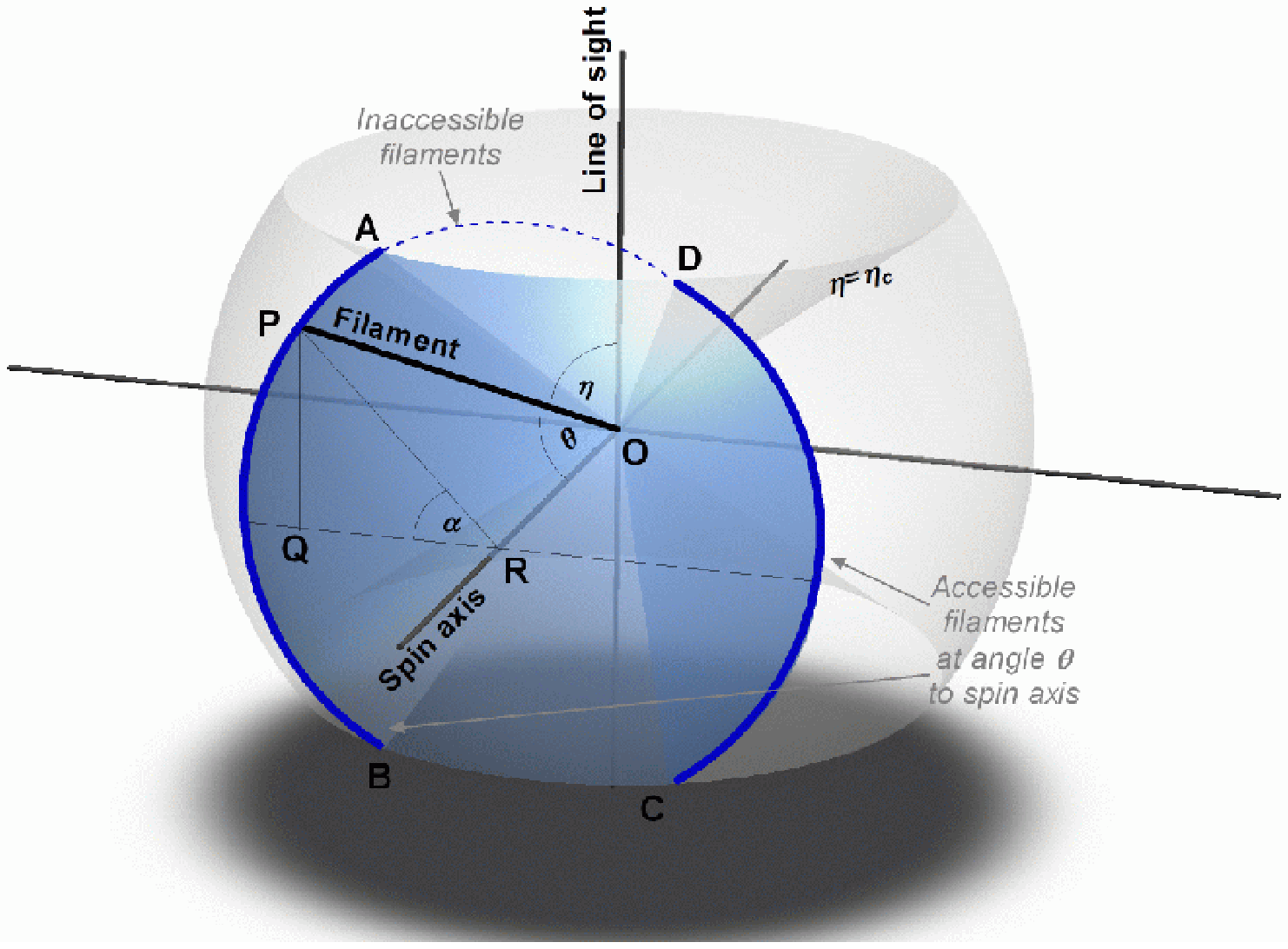} 
  \caption{Distribution of spins in filaments with censored distribution of angles to line of sight.  See also figure \ref{fig:angles}.  The points $A,B,C,D$ lie on the base of a circular cone of semi vertex angle $\theta$.  If there were no censorship of the filament orientation, all angles $\theta$ would be equally accessible.  However, if we eliminate filaments making angles less than $\eta_c$ with the line of sight, the sample of spin orientations $\theta > \pi/2 - \eta_c$ will be truncated.  Only spin axes at angle $\theta$ to the filament lying along the segments $AB$ and $CD$ will be in the sample.  Since the distribution of $\theta$ values around $ABCD$ would be uniform the fraction of objects in the censored sample is the ration of the lengths of the segments $AB$ and $CD$ to the circumference of the circle. }
  \label{fig:censored} 
\end{figure*} 
The angle $\theta$ between the spin vector of a galaxy and its host filament is 
\begin{equation}
\cos \theta = |\cos \Theta \sin \phi|
\end{equation}
(see figure \ref{fig:angles} for the definitions of the angles and equation (\ref{eq:thetadefinition})).  If the filament orientation is random and spins directions are uncorrelated with the host filament, then $\theta$ is randomly distributed over the unit sphere and $\cos \theta$ is uniformly distributed on the interval $[0,1]$.

We have seen that, in our catalogue of filaments, the high-grade Class B and Class B filaments tend to lie transverse to the line of sight (see figure \ref{fig:lineofsight}).  This imposes a constraint on the distribution of angles $\theta$ that the spin axis of a galaxy in our catalogue can make with its host filament.    Even if the filaments and spins were randomly oriented, we would see systematically fewer angles $\theta$ having low values of $\cos \theta$.  The situation is illustrated in figure \ref{fig:censored}.  

\subsection{Definitions}
The geometry of figure \ref{fig:censored} is as follows.  The circle on which the points $A,B,C,D$ lie is the base of a right circular cone with apex at $O$. The side of the cone can be taken to be of unit length.  The axis of the cone, $OR$, is the spin vector of a galaxy in the catalogue.   The point $R$ lies in the plane of the base of the cone, on the axis of the cone.   Thus a point $P$ on this circumference represents a sample filament, and the vector $OP$ makes an angle $\eta$ with the line of sight, and an angle $\theta$ with the spin axis of the galaxy.  $\theta$ is the semi vertex angle of the cone.

The circumference of the base of the cone represents all filaments that are at angle $\theta$ with the spin axis $OR$.  Let us focus attention on the locus of points $P$: the circumference of the base of this cone.  The position of $P$ on the circumference can be parametrised by an angle $\alpha$ in the plane of the base of the cone.  The line $PQ$ in the plane of the base of the cone is parallel to the line of sight, and the angle $\alpha$ is measured from the line $RQ$ (in a clockwise sense).  If everything were randomly and uniformly distributed on a sphere, the distribution of the angle $\alpha$ would be uniform.

\subsection{Censorship}
The angle $\alpha$ is seen to be given in terms of $\eta$ and $\theta$ by the simple relationship
\begin{equation}
\sin \alpha = \frac{\cos \eta}{\sin \theta}
\end{equation}
This follows because $PQ = \cos \eta$ and $PR = \sin \theta$.  From this we have that the values of $\alpha$ traced out as $\eta$ varies for a given $\theta$ are given by
\begin{equation}
\alpha = \arcsin \left[ \frac{\cos \eta}{\sin \theta} \right], \quad \cos \eta < \sin \theta.
\end{equation}
The constraint $\cos \eta < \sin \theta$ shows that not all $\theta$-values are attained at all values of $\eta$.  For a given value of $\theta$ all values of $\alpha$ are attained only if 
\begin{equation}
\eta > \eta_c  \quad \textrm{where} \quad \cos \eta_c = \sin \theta.
\end{equation}
Geometrically, this is depicted in figure \ref{fig:censored} where the base of the cone is shown as the (projected) circle containing the points $A,B,C,D$ and $P$.  

Consider now a constraint such that our catalogue contains only filaments with  $\eta > \eta_c$, for some cutoff $\eta_c$.  The angle $\eta$, as depicted in the figure, is large enough that only points on the segments $AB$ and $CD$ can be reached with values of $\eta$ greater than the censorship angle $\eta_c$.  Points on the segments $AD$ and $BC$ have $\eta < eta_c$, and so the corresponding values of $\alpha$ are unavailable: censorship in $eta$ imposes censorship on $\alpha$.

\subsection{Statistical distribution for random orientations}
\begin{figure}
  \includegraphics[width=0.450\textwidth]{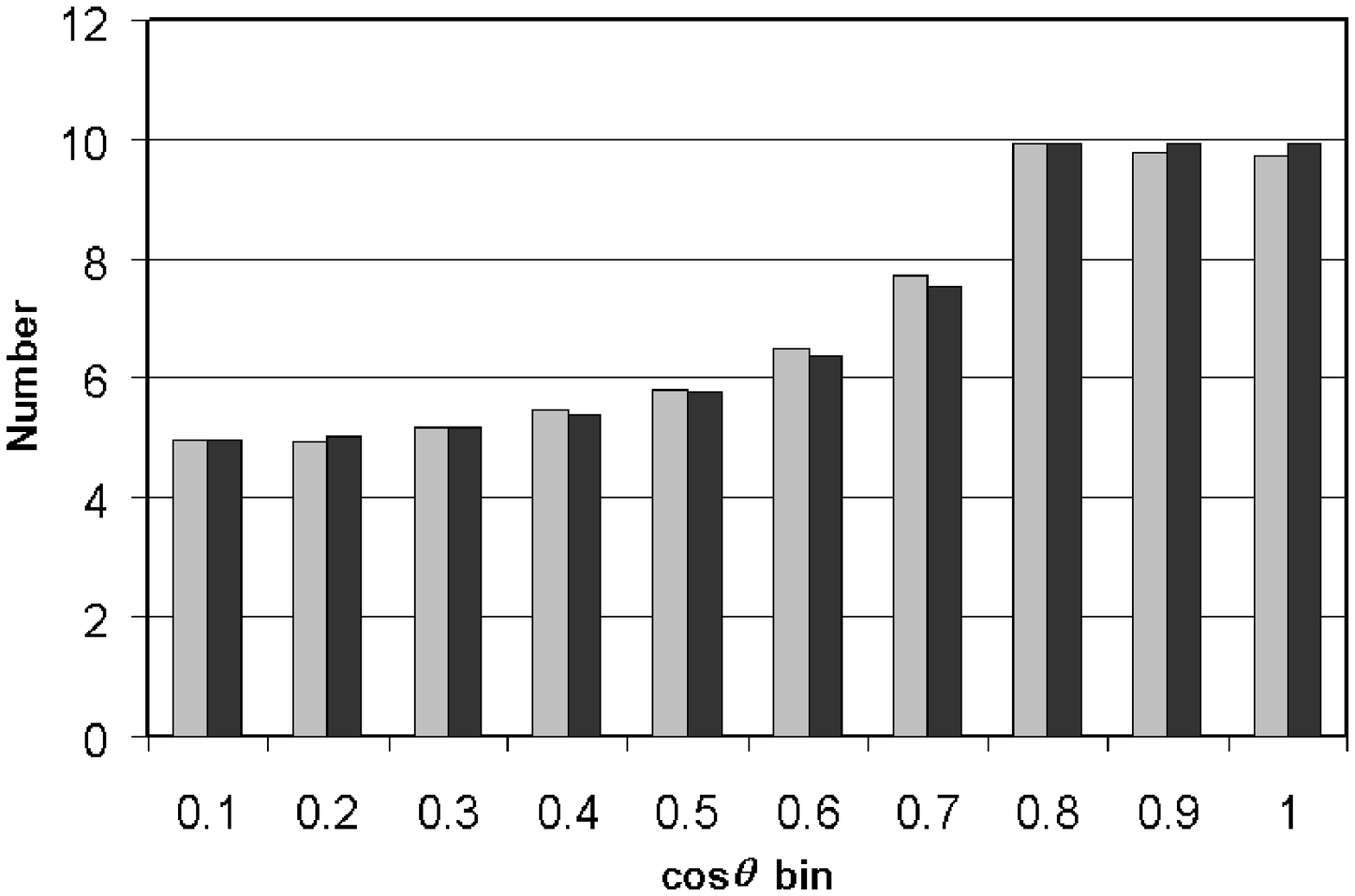} 
  \caption{Comparison of Monte Carlo simulation of randomly oriented galaxies selected from a censored sample of randomly oriented filaments.  The censorship angle is $eta > 45^{\circ}$, as in figure \ref{fig:simulatetheta}.  The Monte Carlo simulation is shown as the light grey histogram.  The dark grey histogram is equation (\ref{eq:select}) appropriately normalised.  }
  \label{fig:MC_model} 
\end{figure} 

In the absence of any $\eta$-censorship the distribution would be uniform in $\cos \theta$.  When a cut-off, $\eta_c$, in $\eta$ is introduced, the uniform distribution is attenuated by a factor
\begin{eqnarray}
f &= \displaystyle \frac{2}{\pi} \alpha = \frac{2}{\pi} \arcsin \left[\frac{\cos \eta_c}{\sin \theta} \right], \quad &\theta > \pi /2 - \eta_c  \nonumber \\
 &= 1 & \textrm{otherwise}.
 \label{eq:select}
\end{eqnarray} 
This can easily be tested using simulations, one of which has been displayed in the grey histogram of figure \ref{fig:simulatetheta}.  The comparison between equation (\ref{eq:select}) and figure \ref{fig:simulatetheta} is shown in figure \ref{fig:MC_model}.

\bsp

\label{lastpage}

\end{document}